%% file: main.tex
\documentclass{article}
\usepackage{graphicx}

\usepackage[left=2.5cm,right=2.5cm,top=2cm,bottom=2.5cm]{geometry}
\usepackage[
    style=nature,
]{biblatex}
\usepackage{float}
\usepackage{color}
\usepackage{multirow}

\usepackage[
    hidelinks,
    colorlinks=false, 
]{hyperref}
\usepackage{authblk}

\input{acronyms}
\input{math-definitions}

\usepackage[font=small,labelfont=bf]{caption}
\usepackage{enumitem}

\crefname{figure}{Figure}{Figures}      

\usetikzlibrary{external}
\begin{document}
\title{Parameter Estimation for Model-Based Sensing of Magneto-Mechanical Resonators}

\author[1,2*]{Sarah Reiss}
\author[1,2,3]{Tobias Knopp}
\author[3]{Justin Ackers}
\author[1,2]{Jonas Faltinath}
\author[1,2]{Fabian Mohn}
\author[1,2]{Marija Boberg}
\author[3]{Nora Timm}
\author[1,2]{Martin Möddel}

\affil[1]{Institute for Biomedical Imaging, Hamburg University of Technology, Hamburg, Germany}
\affil[2]{Section for Biomedical Imaging, University Medical Center Hamburg-Eppendorf, Hamburg, Germany}
\affil[3]{Fraunhofer Research Institution for Individualized Medical Technology and Engineering IMTE, Lübeck, Germany}
\affil[*]{Corresponding author: sarah.reiss@tuhh.de}

\maketitle

\begin{abstract}
\Acp{MMR} represent a recently proposed type of passive sensor that enables the estimation of its pose as well as sensing other parameters in its environment. The working principle of MMRs entails an excitation of the sensors by oscillating magnetic fields, followed by a readout process facilitated by inductive receiver coils. The sensing technology relies on real-time parameter estimation. This encompasses the solution of a nonlinear inverse problem, with the induced signals and a suitable forward model as inputs. The aim of this paper is twofold: first, to introduce a reference model and simplified models for the MMR dynamics and inductive readout, and second, to provide robust and real-time capable methods to estimate the model parameters. The effectiveness of the presented methods is evaluated in terms of their real-time potential, precision, and accuracy. All presented methods demonstrate the capacity to estimate the measured signal, with the simplified methods reducing the corresponding parameter estimation time by up to two orders of magnitude at the expense of less than~\SI{4}{\percent} deviation for large maximum deflection angles.
\end{abstract}

\acresetall

\section{Introduction}  
Parameter estimation is the process of using observed data to infer the values of unknown parameters in a mathematical model, with the aim of representing the underlying system. The forward problem, deriving observed data from a given model, is in general straightforward. However, solving the inverse problem, obtaining a mathematical or physical model that characterizes a series of observed data, is often intricate. With the field of inverse problems being well-established, there are a range of approaches including the least-squares method, the Bayesian approach, and the implementation of neural networks to address this issue~\cite{aster2019parameterestimation, adler2017inverseProbNN}. In practice, measured data are often corrupted by noise and measurement errors, while additional difficulties, including ill-posedness, model uncertainty, and computational complexity, may further complicate parameter estimation. Such challenges arise in a variety of applications, 
from medical imaging for deriving tissue structure to atmospheric remote sensing for acquiring environmental parameters~\cite{mueller2021inverseProblems, wang2012remoteSensing}. 

Online parameter estimation is required in many industrial applications. Unlike offline estimation, where all the data is collected and analyzed afterwards, the parameters have to be acquired and processed in real-time. This capability is essential in domains such as control systems, process monitoring, navigation, and automation, where fast responses to environmental conditions are crucial \cite{zhu2021onlinePMSM, wang2021onlineACC,choudhary2025robotsNN}. Consequently, fast estimation methods are necessary. Possible strategies include using faster estimation methods~\cite{wang2021onlineACC}, predetermining certain parameters using offline estimation algorithms \cite{zhu2021onlinePMSM}, or employing deep-learning approaches \cite{choudhary2025robotsNN}. 

Another technology in which the sensed quantities vary dynamically, and therefore require online parameter estimation, was recently introduced by Gleich et al.~\cite{gleich2023miniature}. They developed a novel tracking and sensing technology based on \acp{MMR}, with possible applications in medical engineering, such as endoscopy, surgery, and vascular interventions~\cite{gleich2023miniature, patentgleich}, as well as process engineering to measure environmental conditions~\cite{merbach2025pressure}. The online parameter estimation is particularly challenging due to measurement noise, errors in the model assumptions, and the computational cost of solving a nonlinear \ac{ODE}. When addressed effectively, the position and orientation of those miniaturized sensors can be acquired in real-time within a large workspace that can, e.g., accommodate the entire human body. Additionally to tracking, \acp{MMR} can be tailored to measure specific environmental parameters, such as temperature~\cite{faltinath2025Temp} or pressure~\cite{merbach2025pressure}, in addition to orientation and position. The technical simplicity of the sensing technology ensures their affordability and facilitates their deployment on a large scale.

In the following, the challenges of operating \acp{MMR} are examined in detail. A complex forward model consisting of a model for the \ac{MMR} dynamics and a model for the measurement system is presented, as well as simplified models derived to overcome the operational difficulties. Additionally, robust and fast methods to estimate the model parameters are presented. A comparative study is conducted to analyze the methods for their real-time capacity, accuracy, and precision using experimental data.

\section{Problem Statement}
\label{sec:problemStatement}
    \begin{figure}[ht!]
        \centering
        \includegraphics[width = \textwidth]{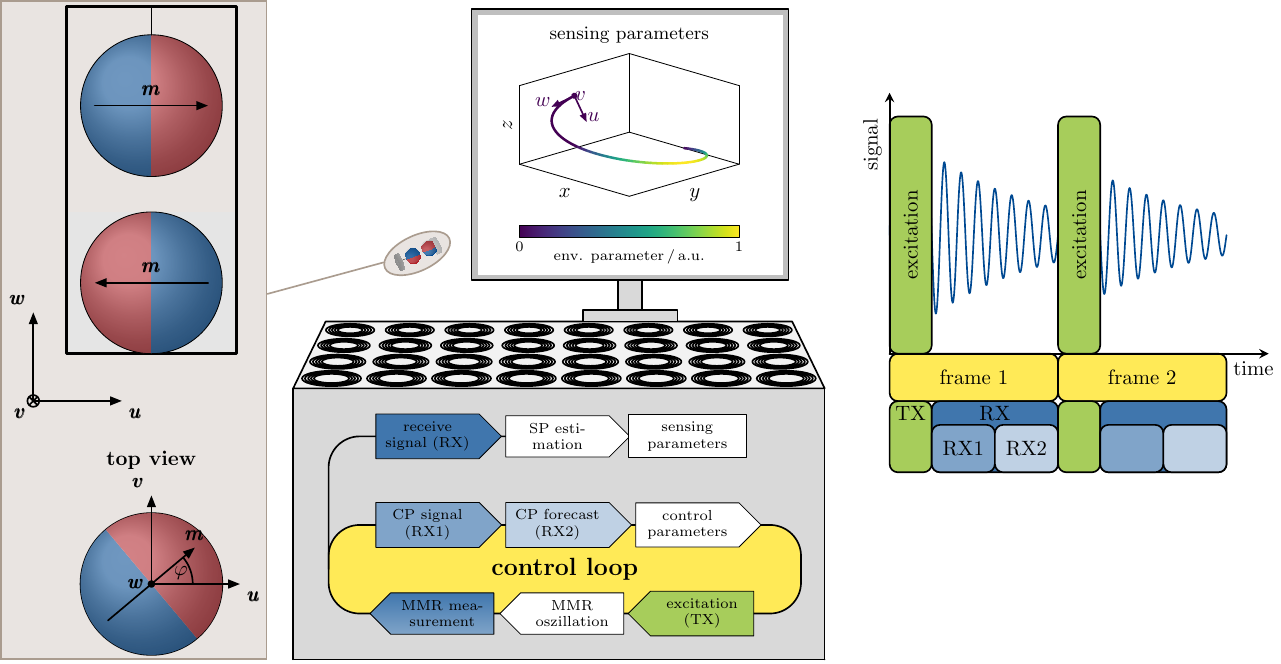} 
        \caption{\textbf{General principle of the \ac{MMR} sensing system.} Left: The \ac{MMR} is composed of two permanent magnets, within a local $uvw$-coordinate system. The lower magnet (stator) is attached to the housing, while the upper one (rotor) is attached to a thin filament shown in black. In the upper left the magnets are in equilibrium alignment, while in the top view of the rotor, the rotor's magnetic moment is displaced by the deflection angle $\varphi$ from its equilibrium position. Center: The schematic representation shows the cyclical flow of measurements and a model experimental configuration for a sensing application. A control loop facilitates continuous operation by establishing control parameters (CP) for the subsequent excitation. The estimation of the sensing parameters (SP) is conducted independently from the control loop on the receive signal. The experimental configuration is equipped with an \ac{MMR} sensor, a sensing system comprising an inductive coil array, and a real-time signal processing platform. Right: An exemplary signal measured by a single coil and the temporal division of the measurement are illustrated. Each frame is divided into an excitation (TX) and receive (RX) window, during which the respective operations are performed.}
        \label{fig:MMRBasics}
    \end{figure}
    
    As described in detail by Gleich et al.~\cite{gleich2023miniature, patentgleich}, the sensing technology consists of three key components: a sensor in the form of an \ac{MMR}, a sensing system based on an inductive coil array employed to excite and read the \ac{MMR} sensor remotely, and a real-time signal processing platform. An \ac{MMR} consists of two magnets. One magnet is affixed to the housing (stator), while the other (rotor) is attached to a thin filament (see \cref{fig:MMRBasics}, left). In their equilibrium position, the magnetic moments are aligned antiparallel, which causes a magnetic attraction that is significantly stronger than the gravitational force and maintains the rotor in a stable suspended state above the stator. The deviation of the rotor's magnetic moment from its equilibrium orientation is quantified by the deflection angle $\varphi$. For deflection angles $-\frac{\pi}{2} < \varphi < \frac{\pi}{2}$, the magnetic torque exerted by the stator drives the rotor into torsional-pendulum-like damped oscillations. 
    
    The sensing system operates in measurement cycles~\cite{gleich2023miniature, patentgleich}(see \cref{fig:MMRBasics}, middle). The term "frame"  denotes an individual cycle, which is further subdivided into two segments. The initial phase is the excitation (TX) window, during which a control signal is transmitted through the coil array to (re-)excite the \ac{MMR} sensor. The second phase is referred to as the receive (RX) window, in which the damped oscillation is read out, typically facilitated by the same coil array. The resulting measurement signal is used twofold: first, for forecasting of the control parameters for the control signal of the following frame, second, for the estimation of the sensing parameters, which are pose and natural frequency.

    Forecasting must be conducted within the hard real-time constraints imposed by the \ac{RX} window. Consequently, the \ac{RX} window is subdivided into an RX1 and an RX2 window (see \cref{fig:MMRBasics}, right). The data acquired in the RX1 window is then utilized to forecast the control parameters during the RX2 window. Accordingly, the efficacy of the forecast method and the duration of RX1 establish an inherent restriction on the duration of  RX2. Therefore, in order to increase the repetition rate of the \ac{MMR} sensing platform, it is possible to decrease the length of the RX1 window. This typically reduces the forecast time, thereby also enabling the shortening of the RX2 window. Alternatively, the forecasting method could be substituted by a faster one. This would also permit a reduction in the RX2 window. However, both of these approaches likely entail a reduction in the quality of the control parameter forecast, so that at a certain point robust excitation of the \ac{MMR} will no longer be feasible. The estimation of the sensing parameters is generally conducted on the signal acquired during the entire receive window (RX1 + RX2). Following the completion of a frame, the latency between signal acquisition and finalization of sensing is caused by the run-time of the estimation method. In circumstances where latency is a concern, a reduction of the input signal and a modification of the estimation method can be employed, albeit at the expense of increased sensing uncertainty.
    
    The forecasting and estimation methods play a pivotal role in achieving optimal performance of the sensing platform, as indicated by repetition rate, latency, and sensing uncertainty. In this study, we analyze a family of model-based methods with respect to the primary contributors to the aforementioned performance indicators: run-time and estimation/forecasting uncertainty.
\section{Methods}

    The methods are structured as follows. First, \cref{sec:timefreqanalysis} introduces an observation-based multi-component model of amplitude- and frequency-modulated sine waves. This model serves as a tool for analyzing the measured signals and obtaining the temporal progression of amplitudes and frequencies. Subsequently, in \cref{sec:signalmodels}, we present a physical model of the MMR experiment and two simplifications thereof. The determination of the model parameters is then the subject of \cref{sec:parameterestimation}. The experimental evaluation and the metrics considered are explained in \cref{subsec:experimentalMethods} and \cref{sec:evaluationmetrics}.

    \subsection{Multi-Component Signal Demodulation} \label{sec:timefreqanalysis}
        The signals observed in \ac{MMR} sensing are caused by the vibratory motion of the sensor. During one period, the magnetic moment of the rotor moves from the equilibrium position to the maximum deflection angle at one end, then to the maximum deflection angle at the other end, and finally back to the equilibrium position. Because the oscillation is damped, the resulting vibration signal is nonstationary, that is, the oscillation frequency and amplitude change over time. The voltage signals observed at the sensing platform are proportional to the time derivatives of the projections of this motion onto the direction specified by the coil sensitivity at the \ac{MMR} position. This signal exhibits the same characteristics as the vibratory motion that caused it.

        We use a multi-component signal of sine waves to model the observed signals. As such
        \begin{equation} 
            s^\text{MC}(t) = \sum_{j=1}^{J} \alpha_j(t)\cos( \psi_j(t)) =  \sum_{j=1}^{J} \left[ \frac{\alpha_j(t)}{2}  e^{\mathrm{i}\psi_j(t)} + \frac{\alpha_j(t)}{2} e^{-\mathrm{i}\psi_j(t)} \right],
            \label{eq:AFMsinewaves}
        \end{equation}
        where the real-valued functions $\alpha_j: \IR^+_0 \rightarrow \IR^+_0$, $\psi_j: \IR^+_0 \rightarrow [0,2\pi)$, and $\omega_j:= \dot{\psi_j}: \IR^+_0 \rightarrow \IR^+_0$ are the instantaneous amplitude, angular phase, and angular frequency of the $j$-th component at time $t$, respectively. We can also express the angular phase as
        \begin{align}
        \psi_j (t)& =\psi_{j,\text{0}}+\int _{0}^{t}\omega_j (\tau )\,\text{d}\tau , \label{eq:integratingPhase} 
        \end{align}
        where $\psi_{j,\text{0}}$ is the initial phase at time $t=0$. Thus, the signal is fully described by $\alpha_j(t)$, $\omega_j(t)$, and $\psi_{j,\text{0}}$.
        
\begin{figure}[t!]
    \centering
    \includegraphics{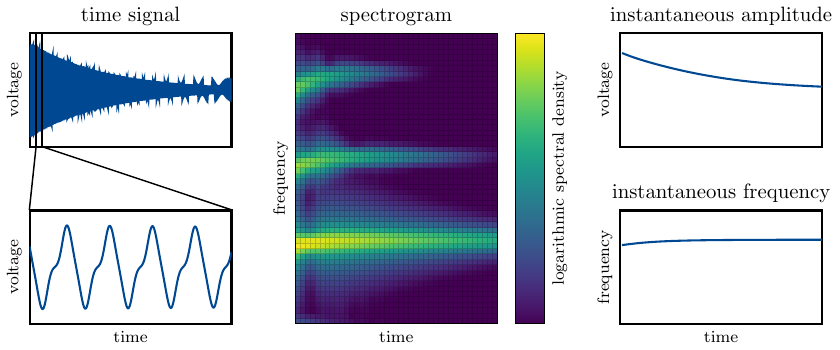} 
    \caption{\textbf{A typical voltage signal detected at the \ac{MMR} sensing platform.} In the time domain, the oscillating nature of the signal is evident in the shorter time frame on the left-hand side. As demonstrated in the time-frequency representation (i.e., the spectrogram $\lvert S_w(t, \tilde{\omega})\rvert^2$ in the center), the signal's multi-component nature is apparent. In analyzing the lowest frequency component, the instantaneous amplitude $\alpha_1(t)$ and frequency $\omega_1(t)$, shown on the right-hand side, are obtained. The former characterizes the signal's decay over time, whereas the latter describes a subtle frequency shift undergone by the signal.}
    \label{fig:Signal}
\end{figure}

        The \ac{STFT} can be used for analyzing amplitude- and frequency-modulated signals. The \ac{STFT} of the signal $s^\text{MC}(t)$ is given by 
        \begin{equation}
             S_w(t, \tilde{\omega}) = \int _{-\infty }^{\infty }s^\text{MC}(\tau)w(\tau-t )\mathrm {e} ^{-\mathrm{i} \tilde{\omega} \tau}\,\mathrm {d} \tau  , \label{eq:STFT}
        \end{equation}
        where $w$ is a window function with support $\text{supp}(w) = [-\frac{1}{2}T_\text{snip},\frac{1}{2}T_\text{snip}]$ of length $T_\text{snip}$ and $\tilde{\omega}$ is the  angular frequency, not to be confused with the actual angular frequencies of the signal $\omega_j(t)$. The spectrogram (see~\cref{fig:Signal}) is defined as the squared magnitude of the \ac{STFT}, 
        usually sampled at a coarse time sampling grid $t_n = (n-1) T_\text{step}$ with step width $T_\text{step}$ for $n=1,\dots,N$, independent of the sampling frequency of the measured signal $s^\text{MC}(t)$. Commonly, the step width is chosen to be less than or equal to the window width, i.e., $T_\text{step} \leq T_\text{snip}$. 
        
        Demodulation of the observed signal into its components defined in \cref{eq:AFMsinewaves} using the STFT
        can be done by assuming that any changes in amplitude and frequency are so small that they remain constant within each bin of the STFT, i.e., for a fixed $t_n$. Thus, we assume
        $\psi_j(t) \approx \omega_{j,n} t + \psi_{j,n}$ and $\alpha_j(t) \approx \alpha_{j,n}$ for $t \in [t_n-\frac{1}{2}T_\text{snip}, t_n+\frac{1}{2}T_\text{snip}]$. As the multi-component sine wave $s^\text{MC}$ in \cref{eq:AFMsinewaves} is a real-valued function, the negative frequency components are redundant. Therefore, we focus on the positive frequency part of the $j$-th component, which is  given by $s^\text{MC}_{j,n}(t) = \frac{1}{2} \alpha_{j,n} \text{e}^{\mathrm{i}( \omega_{j,n} t + \psi_{j,n} )}$. Their corresponding STFT with $W(\tilde{\omega}) = {\cal F}\{ w\} (\tilde{\omega})$ is given by
        \begin{equation}
           S_w^{j,n}(t, \tilde{\omega}) = \frac{\alpha_{j,n}}{2} W(\tilde{\omega} - \omega_{j,n}) \text{e}^{\mathrm{i}((\omega_{j,n} - \tilde{\omega})t + \psi_{j,n})}\label{eq:STFTCompInFS},
        \end{equation}
         see \cref{app:STFTSpectralRepresentation} and Legros et al.~\cite{legros2023}. If we then assume the frequencies of the different components to be sufficiently separated, we can find for each $j$ a neighborhood $\Omega_{j,n} \subset \mathbb{R}$ with $\omega_{j,n} \in \Omega_{j,n}$ such that $S_w(t, \tilde {\omega}) \approx S_w^{j,n}(t, \tilde {\omega})$. Thus, we can use the observation $S_w$ to approximate the individual component $S_w^{j,n}$. Using Equation \cref{eq:STFTCompInFS}, the instantaneous frequency $\hat{\omega}_{j,n}$ and the instantaneous amplitude $\hat{\alpha}_{j,n}$ can then be calculated with
        \begin{equation*} 
             \underset{\hat{\alpha}_{j,n}, \hat{\omega}_{j,n} \in \IR^+_0}{\text{min}}  \int_{-\infty}^{\infty} (|S_w(t, \tilde{\omega})| - \frac{1}{2}\hat{\alpha}_{j,n} W(\hat{\omega}_{j,n} - \tilde{\omega}) )^2 \textrm{d}\tilde{\omega}.
        \end{equation*}
        For the Gaussian window, this can be solved analytically by Gaussian spectrum interpolation~\cite{gasior2004improving}. 
        The initial phase can be directly obtained from \cref{eq:STFTCompInFS} by  $\hat{\psi}_{j,\text{0}} = \text{arg}(S_w(0, \hat{\omega}_{j,1}))$. Since  the temporal resolution of the discrete $\hat{\alpha}_{j,n}$ and $\hat{\omega}_{j,n}$ is constrained by the window width $T_\text{snip}$ and the step width $T_\text{step}$, we use cubic spline interpolation to obtain continuous $\hat{\alpha}_j(t)$ and  $\hat{\omega}_j(t)$. The selection of $T_\text{snip}$ entails a fundamental trade-off: shorter analysis windows provide improved temporal localization at the expense of frequency resolution, whereas longer analysis windows enhance frequency resolution but reduce temporal precision.

In an \ac{MMR} experiment, the measured signal is typically acquired using multiple receive coils, such that the multi-component model \cref{eq:AFMsinewaves} is expressed by a vector-valued signal and amplitude 
\begin{equation} 
    \pmb{s}^\text{MC}(t) = \sum_{j=1}^{J} \pmb{\alpha}_j(t)\cos\!\big(\psi_j(t)\big).    \label{eq:AFMsinewavesmulti}
\end{equation}
In contrast to \cref{eq:AFMsinewaves}, the amplitude vector $\pmb{\alpha}_j: \mathbb{R}^+_0 \rightarrow \mathbb{R}^{L}$ must be allowed to contain signed amplitudes, since the coupling of the \ac{MMR} signal into the receive coils exhibits channel-dependent signs that cannot be absorbed into the scalar cross-channel phase $\psi_j$. Accordingly, after ridge analysis, the instantaneous frequencies $\hat{\omega}_j(t)$, instantaneous signed amplitudes $\hat{\pmb{\alpha}}_j(t)$ and initial phases $\hat{\psi}_{j,\text{0}}$ are obtained. 

\subsection{Signal Models} \label{sec:signalmodels}
We now turn to the corresponding inverse problem: given a model $\pmb{s}^\text{model}(t; \mathcal{P})$ and the measured signal $\pmb{s}^\text{meas}(t)$, the task is to estimate the parameters $\mathcal{P}$ such that
\begin{equation} \label{eq:generalInverseProblem}
     \pmb{s}^\text{meas}(t) \approx\pmb{s}^\text{model}(t; \mathcal{P}), 
    \qquad t \in [0,T],
\end{equation}
where $T$ denotes the length of the considered measurement interval. In the subsequent section, a series of forward models describing the observed signals of a typical \ac{MMR} setup are introduced. First, a base model is derived from physical laws. Second, simplified models are defined using several assumptions. Equation~\cref{eq:generalInverseProblem} is however not limited to the torsional motion models introduced in the following. In principle, more general dynamical models, e.g., incorporating pendulum modes, can be employed, which would naturally involve a different parameter set $\mathcal{P}$.

\begin{figure}[t!]
            \centering
            \includegraphics{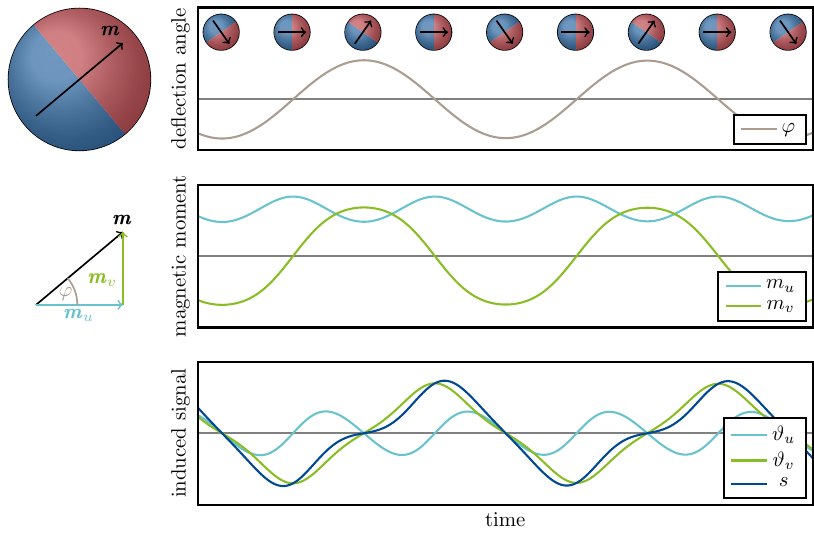} 
            \caption{\textbf{Basic principle of signal generation.} Top: Within the receive window, the magnetic moment of the rotor oscillates freely. Its deflection angle $\varphi$ from its equilibrium position $\varphi = 0$ and corresponding orientation are shown. Center: The movement causes two fundamental oscillation frequencies of the magnetic moment in the $uvw$-coordinate system: the $v$-component, which oscillates at the same frequency around zero, and the $u$-component, which oscillates with twice the frequency around a nonzero offset. Bottom: If we align a unit sensitivity sensing system to pick up the time derivative of each of the individual components, we would observe $\vartheta_u$ and $\vartheta_v$, respectively. A typical signal will be a superposition of these, such as $s$, shown in the same plot.}
            \label{fig:baseSignal}
        \end{figure}

\subsubsection{Signal Equation}

We consider a sensing system based on an inductive coil array using $L$ receive coils that are characterized by their receive coil sensitivity $\pmb{P}: \mathbb{R}^3 \rightarrow \mathbb{R}^{L \times 3}$ that takes the position as the input and outputs a matrix containing the receive coil sensitivities as rows \cite{jin2018electromagnetic}. The receive coil sensitivity is formally defined as the magnetic flux density that an individual coil would generate per unit current. We further consider only the receive window in the absence of an external magnetic (excitation) field. 

The \ac{MMR}'s pose is characterized by the position of its rotor $\pmb{r} \in \mathbb{R}^3$ in space and its orientation that can be described by the $3\times 3$ rotation matrix $\pmb{R} \in SO(3)$. We assume that both are static during one frame (quasi-static assumption), which in practice implies that the measurement frame rate needs to be high enough that the movement of the \ac{MMR} during one frame is negligible. The magnetic moment of the rotor $\pmb{m}: [0,T] \rightarrow \mathbb{R}^3$ in the local $uvw$-coordinate system of the \ac{MMR} (see \cref{fig:MMRBasics}) changes due to the vibration of the \ac{MMR} and can be expressed in the global $xyz$-coordinate system as
$\pmb{m}_\text{global}(t) = \pmb{R}\, \pmb{m}(t)$. Based on Faraday's induction law, we can then express the voltage signals $\pmb{s}: [0,T] \rightarrow \mathbb{R}^L$ induced in the receive coils during a single frame as
\begin{align} \label{Eq:SignalEquation}
    \pmb{s}(t; \pmb{r}, \pmb{R}) & = -\pmb{P}(\pmb{r}) \,  \partial_t \pmb{m}_\text{global}(t) =  -\pmb{P}(\pmb{r}) \pmb{R}\, \partial_t \pmb{m}(t).
\end{align}
We can combine the coil sensitivity $\pmb{P}$ and the rotation matrix $\pmb{R}$ into the signal projection matrix 
\begin{align}
  \pmb{\sigma}(\pmb{r}, \pmb{R}) &:= -\pmb{P}(\pmb{r}) \pmb{R}  
  \label{eq:signalProj}
\end{align}
yielding 
\begin{align} \label{Eq:SignalEquation2}
    \pmb{s}(t; \pmb{r}, \pmb{R}) & =  \pmb{\sigma}(\pmb{r}, \pmb{R}) \, \partial_t \pmb{m}(t).
\end{align}
Thus, the signal projection matrix determines the weights by which the $uvw$-components of the derivative of the magnetic moment combine into the individual receive channels.

The general task within an \ac{MMR} experiment is to take the measured voltage signals $\pmb{s}(t)$ and estimate the unknown parameters $\pmb{r}$, $\pmb{R}$, and $\pmb{m}(t)$, a nonlinear inverse problem that is challenging to solve. However, the task can be divided into different subtasks that are easier to handle. Effective control and sensing of an environmental parameter requires an understanding of the MMR’s dynamic behavior and an estimation of its magnetic moment $\pmb{m}(t)$. In both cases, it is not strictly necessary to determine the \ac{MMR}'s position $\pmb{r}$ and orientation $\pmb{R}$; instead, the signal projection matrix $\pmb{\sigma}$ can be estimated as an intermediate representation. The focus of this work is an accurate estimation of \cref{Eq:SignalEquation2}. Subsequently, the position $\pmb{r}$ and orientation $\pmb{R}$ can be extracted from $\pmb{\sigma}$ solving \cref{eq:signalProj}.

One important assumption that is commonly made for well-designed \acp{MMR} is that the rotor is strictly performing a 1D torsion around the $w$ axis, as shown in~\cref{fig:MMRBasics}. The derivative of the magnetic moment in the $uvw$-coordinate system can then be expressed as
\begin{equation} \label{Eq:mderivsep}
\partial_t \pmb{m}(t) = m_\text{r} \begin{pmatrix} \partial_t \cos(\varphi(t)) \\ \partial_t\sin(\varphi(t)) \\ 0
\end{pmatrix} = m_\text{r} \begin{pmatrix} \theta_u(t) \\ \theta_v(t) \\  0 \end{pmatrix},
\end{equation}
where $m_\text{r}$ is the magnitude of the magnetic moment vector of the rotor,  $\varphi: [0,T] \rightarrow (-\frac{\pi}{2},\frac{\pi}{2})$ is the deflection angle within the $uv$-plane with respect to the $u$-axis, and
\begin{align}
\theta_u(t) & :=  \partial_t \cos(\varphi(t)) = -  \dot{\varphi}(t) \sin(\varphi(t)) \label{Eq:templateU}\\
\theta_v(t) & :=  \partial_t \sin(\varphi(t))  =   \dot{\varphi}(t) \cos(\varphi(t)) \label{Eq:templateV}
\end{align}
are so-called basis functions, which describe the local sensor signal  within the $uvw$-coordinate system. 
With this, we can express \cref{Eq:SignalEquation2} as
\begin{align} \label{Eq:SignalEquation3}
    \pmb{s}(t) & =  \pmb{\sigma} \, \partial_t \pmb{m}(t) = m_\text{r}(\pmb{\sigma}_v \theta_v(t) + \pmb{\sigma}_u \theta_u(t)  ),
\end{align}
where $\pmb{\sigma}_u$ is the first and $\pmb{\sigma}_v$ is the second column of $\pmb{\sigma}$.
Thus, each channel contains a mixture of the $u$-signal $\theta_u(t)$ and the $v$-signal $\theta_v(t)$ with the signal projections $\pmb{\sigma}_u$ and $\pmb{\sigma}_v$ acting as the mixing factors. While $\theta_v(t)$ shows oscillating behavior with an instantaneous frequency $\omega$ and its odd harmonics (due to the sine projection of the deflection angle $\varphi$ in \cref{Eq:mderivsep}), $\theta_u(t)$ contains a signal at $2\omega$ and higher even harmonics. \cref{fig:baseSignal} shows typical \ac{MMR} signals and the relations among $\pmb{s}(t)$, $\varphi(t)$, $\theta_v(t)$, and $\theta_u(t)$. 

Both the magnetic moment $m_\text{r}$ and the two projection vectors $\pmb{\sigma}_v$ and $\pmb{\sigma}_u$ enter linearly into \cref{Eq:SignalEquation3}. Consequently, these parameters are not uniquely identifiable, since scaling factors can be redistributed between $m_\text{r}$ and the projection vectors without altering the predicted signal. To resolve this ambiguity, we fix $m_\text{r}$ as a known constant throughout this paper. In practice, this means that the magnetic moment $m_\text{r}$  needs to be acquired once for each sensor (see \cite{faltinath2025natural}), and can then be used during the sensor's lifetime. 

\subsubsection{Torsion Model (T)}

Assuming the rotor has only one torsional degree of freedom, Gleich et al.~\cite[Supplemental Material]{gleich2023miniature} showed that the temporal evolution of the deflection angle $\varphi$ during RX is governed by the nonlinear ordinary differential equation
\begin{equation} \label{Eq:DeflAngleODE}
\ddot{\varphi} + \frac{\omega_\text{nat}}{Q} \dot{\varphi} + \omega_\text{nat}^2 \sin{\varphi} = 0.
\end{equation}
Here, the quality factor $Q$ characterizes damping; a high value ensures that the dissipative contribution, the $\dot{\varphi}$ term, remains small. The natural frequency of the resonator
\begin{equation*}
\omega_\mathrm{nat} = \sqrt{\frac{m_\text{r}B}{I}}
\end{equation*}
depends on the moment of inertia $I$, 
the magnitude of the magnetic moment vector $m_\text{r}$ of the rotor, and the external magnetic field $B$ caused by the stator generating the torque. The natural frequency can be used to encode an environmental parameter. 

Equation~\cref{Eq:DeflAngleODE} is a nonlinear second-order \ac{ODE} with initial conditions $\varphi(0)$ and $\dot{\varphi}(0)$. It describes an anharmonic oscillator analogous to a gravitational pendulum. While no closed-form analytic solution exists, numerical integration is straightforward and was used for the simulations illustrating the fundamentals of signal generation in \cref{fig:baseSignal}. Qualitatively, the solution is oscillatory with an instantaneous frequency $\omega$ that satisfies $\omega < \omega_\text{nat}$ and approaches $\omega_\text{nat}$ in the limit $\varphi \to 0$ and $Q \rightarrow \infty$. Thus, the oscillation frequency increases as the deflection decreases. At the same time, the amplitude decays approximately exponentially. Only in the case of small deflection angles (see \cref{subsec:smallangle}), the amplitude decays exactly exponentially at the rate
\begin{equation}
\label{eq:lambdaQ}
    \lambda := \frac{\omega_\text{nat}}{2Q}.
\end{equation}
The relaxation parameter $\lambda$ is selected as the sole model parameter for damping, as the quality factor can be directly obtained from a given natural frequency.

Instead of directly specifying the initial values, it is convenient to parameterize the model by the maximum deflection angle $\varphi_\text{max}$, the maximum angular velocity $\dot{\varphi}_\text{max}$, and an initial phase $\psi_\text{start}$, such that
\begin{align}
      \varphi(0) &= \varphi_\text{max} \cos(\psi_\text{start}), \\
      \dot{\varphi}(0) &= -\dot{\varphi}_\text{max} \sin(\psi_\text{start}).
\end{align}
Note that maximum deflection angle and maximum angular velocity are related via
\begin{equation}
\label{eq:velocitydeflection}
\dot{\varphi}_\text{max} = \,\omega_\text{nat}\,\sqrt{2-2\cos(\varphi_\text{max})},
\end{equation}
as shown in \cref{app:velocitydeflection}. The torsion model then yields the induced signal 
\begin{align} \label{Eq:SignalEquationTorsionModel}
    \pmb{s}^\text{T}(t; \mathcal{P}) & = m_\text{r}(\pmb{\sigma}_v \dot{\varphi}(t) \cos(\varphi(t)) - \pmb{\sigma}_u \dot{\varphi}(t) \sin(\varphi(t))),
\end{align}
which is fully characterized by the parameters $\mathcal{P} := \{ \pmb{\sigma}_u, \pmb{\sigma}_v, \omega_\text{nat}, \lambda, \varphi_\text{max}, \psi_\text{start} \}$.

In the following sections, we introduce two independent simplifications leading to analytic solutions of \cref{Eq:DeflAngleODE}: the small-angle torsion model (TSA) and the undamped torsion model (TU). However, neither model captures all relevant features of the MMR signal. The TSA does not account for the amplitude-dependent frequency shift, whereas the TU neglects the damping. In \cref{subsec:TUSA}, we therefore combine both models into the undamped and small-angle torsion model (TUSA).

\subsubsection{Small-Angle Torsion Model (TSA)}
 \label{subsec:smallangle}

In the small-angle model, it is assumed that $\sin \varphi \approx \varphi$, which holds for sufficiently small deflection angles (typically below 14$^\circ$). This reduces the equation of motion~\cref{Eq:DeflAngleODE} to that of a damped harmonic oscillator,
\begin{equation} \label{Eq:HarmonicOscillator}
\ddot{\varphi} + \frac{\omega_\text{nat}}{Q} \dot{\varphi} + \omega_\text{nat}^2 \varphi = 0,
\end{equation}
which has the analytic solution \cite{lelas2023damped}
\begin{equation}
\varphi^\text{TSA}(t) := \varphi_\text{env}(t) \cos \left(\sqrt{1-\left(\tfrac{1}{2Q}\right)^2}\, \omega_\text{nat}t+\psi_\text{start} \right), 
\qquad
\varphi_\text{env}(t) := \varphi_\text{max} \text{e}^{-\lambda t}, \label{eq:expDecay}
\end{equation}
where $\varphi_\text{env}(t)$ is the envelope function of the oscillating $\varphi$.

This solution captures two essential features of numerical solutions to Equation \cref{Eq:DeflAngleODE}: the exponential decay and a frequency that is slightly lower than $\omega_\text{nat}$ for high $Q$. However, it fails to describe the amplitude-dependent frequency shift of the \ac{MMR} signal.

Using the small-angle torsion model, the induced signal can be then expressed as 
\begin{align} \label{Eq:SignalEquationTSAModel}
    \pmb{s}^\text{TSA}(t; \mathcal{P}) & = m_\text{r}(\pmb{\sigma}_v \dot{\varphi}(t) \cos(\varphi(t)) - \pmb{\sigma}_u \dot{\varphi}(t) \sin(\varphi(t))),
\end{align}
with the same parameters as the torsion model, namely $\mathcal{P} = \{ \pmb{\sigma}_u, \pmb{\sigma}_v, \omega_\text{nat}, \lambda, \varphi_\text{max}, \psi_\text{start} \}$.

\subsubsection{Undamped Torsion Model (TU)}

Another simplification is to neglect damping ($Q \rightarrow \infty$), yielding the undamped model
\begin{equation} \label{Eq:DeflAngleODEUndamped}
\ddot{\varphi} + \omega_\text{nat}^2 \sin{\varphi} = 0.
\end{equation}
Its solutions are strictly periodic without amplitude decay. The oscillation frequency depends on the maximum deflection angle and is given by
\begin{equation}
    \omega^\text{TU}(\varphi_\text{max}) := g(\varphi_\text{max})\, \omega_\text{nat}, 
    \qquad 
    g(\varphi_\text{max}) := \frac{\pi}{2K(\sin (\varphi_\text{max}/2))},  \label{Eq:omegaUndampedFormula}
\end{equation}
where $K(k) = \int_{0}^{\pi/2} \frac{1}{\sqrt{1-k^2\sin^2 \Psi}} \,\mathrm{d}\Psi$ denotes the complete elliptic integral of the first kind~\cite{belendez2007solNonlinearPend}. 
The frequency $\omega^\text{TU}(\varphi_\text{max})$ is always smaller than $\omega_\text{nat}$. The crucial difference from the small-angle torsion model is that the undamped torsion model accounts for the decrease in frequency with increasing amplitude.

An explicit analytical solution to \cref{Eq:DeflAngleODEUndamped} can be expressed as a Fourier series (see \cref{app:Fourierseriesderivation})
\begin{align}
   \varphi^\text{TU}(t) & := 8 \sum_{n=0}^{\infty} c_n(\varphi_\text{max}) \cos\!\left((2n+1)\,\omega^\text{TU}(\varphi_\text{max}) t + \psi_\text{start}\right), \label{Eq:phiSeries}\\
   c_n(\varphi_\text{max}) & = \frac{(-1)^n}{2n+1}\, \frac{\big(q(\varphi_\text{max})\big)^{n+1/2}}{1+\big(q(\varphi_\text{max})\big)^{2n+1}}, 
   \qquad 
   q(\varphi_\text{max}) := \exp \!\left( -\pi \frac{K\left(\sqrt{1-\sin^2 (\varphi_\text{max}/2)}\right)}{K(\sin (\varphi_\text{max}/2))} \right). \label{Eq:phiFT}
\end{align}
The Fourier coefficients $c_n$ decay rapidly, so that for small amplitudes $\varphi_\text{max}$ the motion is well approximated by the first term, i.e., by a purely sinusoidal oscillation. For larger amplitudes, however, higher harmonics become significant and visibly distort the waveform.

Assuming the undamped torsion model with $c_n = 0$ for $n > 0$, we show in \cref{sec:appDerivationFourier} that the induced signal can be expressed as
\begin{align}
     \pmb{s}^\text{TU}(t;\mathcal{P} \setminus \{ \lambda \} ) 
    &:= \pmb{\alpha}_v\!\big(\varphi_\text{max}\big)\,
    \sin\!\big(\psi^\text{TU}(t)\big) 
    + \pmb{\alpha}_u\!\big(\varphi_\text{max}\big)\,
    \sin\!\big(2\psi^\text{TU}(t) \big), 
    \label{Eq:signalUndampedExp}
\end{align}
where the amplitude functions are given by
\begin{align}
    \pmb{\alpha}_v(\varphi_\text{max} )
    &:= -m_\text{r}\,\kappa_v(\varphi_\text{max})\,\omega_\text{nat}\,\pmb{\sigma}_{v},
    \quad 
    \kappa_v(\varphi_\text{max}) := 
    \frac{\pi^3}{K\!\big(\sin (\varphi_\text{max}/2)\big)^3}\,
    \frac{\sqrt{q(\varphi_\text{max})}}{1+q(\varphi_\text{max})},
    \label{Eq:AvUndamped}\\[0.5em]
    \pmb{\alpha}_u(\varphi_\text{max}) 
    &:= 4m_\text{r}\,\kappa_u(\varphi_\text{max})\,\omega_\text{nat}\,\pmb{\sigma}_{u},
    \quad     
    \kappa_u(\varphi_\text{max}) := 
    \frac{\pi^3}{K\!\big(\sin (\varphi_\text{max}/2)\big)^3}\,
    \frac{q(\varphi_\text{max})}{(1+q(\varphi_\text{max}))^2}
    \label{Eq:AuUndamped}
\end{align}
and the angular phase by $\psi^\text{TU}(t) = \omega^\text{TU}(\varphi_\text{max}) t+\psi_\text{start}$.

The undamped torsion model is parameterized by $\mathcal{P} \setminus \{ \lambda \} = \{ \pmb{\sigma}_u, \pmb{\sigma}_v, \omega_\text{nat},  \varphi_\text{max}, \psi_\text{start} \}$, i.e., compared to the torsion model, the relaxation term $\lambda$ is absent.

\subsubsection{Combined Undamped and Small-Angle Torsion Model (TUSA)}
\label{subsec:TUSA}
As both models lack essential features, we propose coupling the two models. The combined undamped and small-angle torsion model (TUSA) uses the undamped torsion model \cref{Eq:signalUndampedExp} as the basis. It then replaces the static deflection angle $\varphi_\text{max}$ by the exponentially decaying deflection angle $\varphi_\text{max}^\text{TSA}(t)$ 
from the small-angle model \cref{eq:expDecay} and replaces the angular phase $\psi^\text{TU}(t)$ with 
\begin{equation} \label{eq:angularPhasePhiMax}
    \psi^\text{TUSA}(t) := \psi_\text{start} + \int_{0}^{t} 
    \omega^\text{TU}\!\big(\varphi_\text{env}(\tau)\big)\, \mathrm{d}\tau,
\end{equation}
which is based on \cref{eq:integratingPhase}. In combination, this yields
\begin{align}
    \pmb{s}^\text{TUSA}(t;\mathcal{P}) 
    & = \pmb{\alpha}_v\!\big(\varphi_\text{env}(t)\big)\,
    \sin\!\big(\psi^\text{TUSA}(t)\big) 
    + \pmb{\alpha}_u\!\big(\varphi_\text{env}(t)\big)\,
    \sin\!\big(2 \psi^\text{TUSA}(t) \big),
\end{align}
with $\pmb{\alpha}_v$ and $\pmb{\alpha}_u$ as defined in \cref{Eq:AvUndamped} and \cref{Eq:AuUndamped}. The TUSA model is parameterized by the complete parameter set $\mathcal{P}$, i.e., it has the same parameters as the torsion model. 

A comparison with the multi-component sine wave model $\pmb{s}^\text{MC}(t)$ in \cref{eq:AFMsinewavesmulti} reveals that the TUSA model $\pmb{s}^\text{TUSA}(t;\mathcal{P})$ is a special case of the MC model. This can be shown by considering the following correspondences:
\begin{align}
    \pmb{\alpha}_1(t) \,&\hat{=}\,\pmb{\alpha}_v\!\big(\varphi_\text{env}(t)\big), \quad \psi_1(t) \,\hat{=}\, \psi^\text{TUSA}(t) - \frac{\pi}{2},\label{eq:MCeqTUSA0}\\
    \pmb{\alpha}_2(t) \,&\hat{=}\,\pmb{\alpha}_u\!\big(\varphi_\text{env}(t)\big), \quad \psi_2(t) \,\hat{=}\, 2\psi^\text{TUSA}(t) - \frac{\pi}{2}.
\label{eq:MCeqTUSA1}
\end{align}
The phase shift $-\frac{\pi}{2}$ is necessary because $\pmb{s}^\text{TUSA}(t)$ is formulated using sine functions, while $\pmb{s}^\text{MC}(t)$ uses cosine functions. This relationship is justified by the trigonometric identity $\cos\left(x-\frac{\pi}{2}\right) = \sin(x)$.

The ability to express the TUSA model in the form of the MC model is important because it permits the use of time-frequency analysis to derive the parameters of the TUSA model, as discussed in \cref{subsec:timeFreqMethods}.

\subsection{Parameter Estimation} \label{sec:parameterestimation}
In the preceding section, we derived the distinct forward models $\pmb{s}^\text{T}$ and $\pmb{s}^\text{TUSA}$ for computing the induced signals of an \ac{MMR} experiment as functions of the model parameters $\mathcal{P}$. 

\begin{figure}
\centering
    \includegraphics{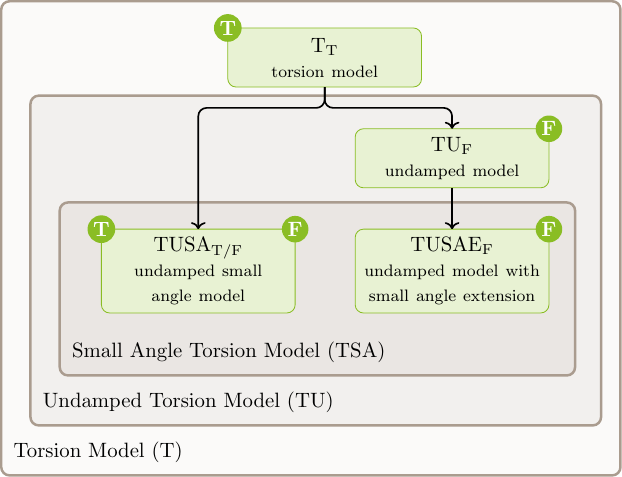}
    \caption{\textbf{Overview of the models and methods.} The abbreviations T (time) and F (time-frequency) indicate the corresponding domain for the parameter estimation method.}
    \label{fig:overviewModels}
\end{figure}

In the following, we discuss different strategies to obtain the parameter set $\mathcal{P}$ by solving the inverse problem \cref{eq:generalInverseProblem}, either directly in the time domain on the measured signal, or based on the instantaneous functions from the time-frequency domain. \cref{fig:overviewModels} gives an overview of the proposed methods. We use a naming scheme for the method that is composed of the model and fitting domain considered, i.e., Model$_\text{Domain}$. For the model, we consider the torsion model (T), the undamped model (TU), and the combined undamped small-angle model (TUSA). We also consider a model that primarily assumes the undamped model but uses the small-angle approximation to fit remaining parameters ($\varphi_\text{max}$, $\lambda$) that are inaccessible with the undamped model, named undamped model with small-angle extension (TUSAE). For the fitting domain, we consider time domain (T), using $\pmb{s}^\text{meas}(t)$ directly, and time-frequency domain (F), using the instantaneous amplitude and/or frequency acquired with the STFT.

\subsubsection{Time-Based Methods}

One straightforward way to solve \cref{eq:generalInverseProblem} is to minimize the model deviation in the time domain considering the box-constraint nonlinear least-squares problem
\begin{equation} \label{Eq:ParamIdentGeneral}
     \underset{\mathcal{P} \in \Gamma}{\text{min}}  \int_{0}^{T} \Vert \pmb{s}^\text{meas}(t) - \pmb{s}^\text{model}(t ; \mathcal{P})\Vert_{\breve{\pmb{W}}(t)}^2 \,\text{d}t .
\end{equation}
Here, $\breve{\pmb{W}}(t)$ is a diagonal weighting matrix function that allows to control the influence of different channels at time $t$. Problem \cref{Eq:ParamIdentGeneral} needs to be solved in an iterative fashion using a suitable optimization algorithm. The cost function can have many local minima such that convergence to the actual solution cannot be guaranteed and, in fact, often depends on the initial parameters. We present a suitable optimizer in \cref{sec:optimization}. $\Gamma$ are constraints that restrict the solution space. Suitable initial parameters and the selection of $\Gamma$ are discussed in \cref{sec:startAndConstraints}. A proper selection enables incorporation of prior knowledge. 

We call the time-based method that numerically solves the \ac{ODE} \cref{Eq:DeflAngleODE} T$_\text{T}$ (torsion model in time domain) and the one that uses the simplifications TUSA$_\text{T}$ (undamped small-angle model in time domain). 

\subsubsection{Time-Frequency-Based Methods}
\label{subsec:timeFreqMethods}

While the time-based optimization method \cref{Eq:ParamIdentGeneral} is straightforward to implement, it always requires evaluations of the full time series. Considering the simplified models, it is possible to solve the problem in the time-frequency domain, which drastically reduces the computational effort.

To this end, we first perform a time-frequency analysis on $\pmb{s}^\text{meas}$ using the STFT method discussed in \cref{sec:timefreqanalysis}. We obtain the functions $\hat{\omega}(t)$, $\hat{\pmb{\alpha}}_v(t)$, and $\hat{\pmb{\alpha}}_u(t)$, which serve as the basis for determining the model parameters of $\mathcal{P}$. In addition, we exploit the direct relation \cref{eq:MCeqTUSA0} between the MC model and the TUSA model, which provides an immediate estimate of the initial phase $\hat{\psi}_\text{start}$. In the following, we discuss several algorithmic strategies for this task.

\paragraph{Method TUSA$_\text{F}$ (combined undamped and small-angle torsion model in time-frequency domain):}  
In the first approach, the estimates of the instantaneous functions $\hat{\omega}(t)$, $\hat{\pmb{\alpha}}_v(t)$, and $\hat{\pmb{\alpha}}_u(t)$ are linked to the model of the instantaneous functions $\omega^\text{TU}( \varphi_\text{env}(t))$, $\pmb{\alpha}_v(\varphi_\text{env}(t))$, and $\pmb{\alpha}_u(\varphi_\text{env}(t))$ considering the nonlinear weighted least-squares problem
\begin{align}
   \underset{ \mathcal{P} \setminus \{ \psi_\text{start} \} \in \Gamma \setminus \{ \psi_\text{start} \} }{\text{min}} \int_{0}^T \left(\breve{w}(t) \left( \hat{\omega}(t) - \omega^\text{TU}( \varphi_\text{env}(t)) \right)^2 + \sum_{\iota \in \{v,u\}}\Vert  
   \hat{{\pmb{\alpha}}}_\iota(t) - {\pmb{\alpha}}_\iota(\varphi_\text{env}(t))  \Vert_{\breve{\pmb{W}}_\iota(t)}^2 \right) \text{d}t . \label{Eq:OptModelD}
\end{align}
Here, $\breve{w}(t)$, $\breve{\pmb{W}}_v(t)$, and $\breve{\pmb{W}}_u(t)$ are weights that can be used to change the influence of different time points and channels on the cost function. The norm is the Frobenius norm. 

\paragraph{Method TU$_\text{F}$ (undamped torsion model in time-frequency domain):} In method TUSA$_\text{F}$, the instantaneous amplitude and frequency data is connected through the exponential decay model $\varphi_\text{env}(t)$. The following section describes method TU$_\text{F}$, which does not require modeling of the envelope function because it removes the explicit dependency on the deflection angle. The auxiliary function $g(\varphi_\text{max})$ in the expression for the frequency \cref{Eq:omegaUndampedFormula} as well as $\kappa_v(\varphi_\text{max})$ and $\kappa_u(\varphi_\text{max})$ in the expressions for the amplitudes \cref{Eq:AvUndamped} and \cref{Eq:AuUndamped} only depend on the maximum deflection angle. To exploit this fact, we can invert the monotonic function $\kappa_v$ and $\kappa_u$ numerically by calculating a dense look-up table and interpolation. 

After insertion of $\pmb{\alpha}_v(t)$ and $\pmb{\alpha}_u(t)$, the Equations \cref{Eq:AvUndamped} and \cref{Eq:AuUndamped} can be rearranged, and inserted into \cref{Eq:omegaUndampedFormula}
\begin{equation*}
\begin{aligned}
    \omega^\text{TU}_{v}( \omega_\text{nat}, \sigma_{v,l}, \alpha_{v,l}(t)) &= g\left( \kappa_v^{-1} \left( -\frac{\alpha_{v,l}(t)}{\omega_\text{nat} \sigma_{v,l}} \right) \right) \omega_\text{nat},\\
  \omega^\text{TU}_{u}(\omega_\text{nat}, \sigma_{u,l}, \alpha_{u,l}(t)) &= g\left( \kappa_u^{-1} \left( \frac{\alpha_{u,l}(t)}{4\omega_\text{nat}\sigma_{u,l}} \right) \right) \omega_\text{nat}.     
\end{aligned}
\end{equation*}

Hence, we eliminated the explicit dependency on the maximum deflection angle and do not require any assumption on its progression over time.

Using the instantaneous frequency  $\hat{\omega}(t)$ and amplitudes $\hat{\alpha}_{v,l}(t)$ and $\hat{\alpha}_{u,l}(t)$, we can then formulate the nonlinear weighted least-squares problem
\begin{equation} \label{Eq:OptModelC}
   \underset{ \mathcal{P} \setminus \{ \lambda, \varphi_\text{max}, \psi_\text{start} \} \in \Gamma \setminus \{ \lambda, \varphi_\text{max}, \psi_\text{start} \}}{\text{min}} 
  \sum_{\iota \in \{u,v\}} \sum_{l=1}^{L} \int_{0}^T \breve{w}_{\iota,l}(t) \left(  \hat{\omega}(t) - \omega^\text{TU}_{\iota}(\omega_\text{nat}, \sigma_{\iota,l}, \hat{\alpha}_{\iota,l}(t))  \right)^2 \text{d}t  ,
\end{equation}
where $\breve{w}_{\iota,l}(t)$ are weights that can be used to emphasize certain data (e.g., the receive channel with the highest signal). 

\paragraph{Method TUSAE$_\text{F}$ (combined undamped torsion model with small-angle extension in time-frequency domain):} 

Method TU$_\text{F}$ is not able to estimate the maximum deflection angle $\varphi_\text{max}$ nor the relaxation parameter $\lambda$. However, an estimate of the envelope function $\hat{\varphi}_{\text{env}}(t)$ can be obtained by inserting the solution from \cref{Eq:OptModelC} and the instantaneous amplitude $\hat{\alpha}_{v,l}(t)$ or $\hat{\alpha}_{u,l}(t)$ into the inversion of \cref{Eq:AvUndamped} or \cref{Eq:AuUndamped} 
\begin{align*}       \label{Eq:TU_inversionAmp}
     \hat{\varphi}_{\text{env}}(t) & = \kappa_v^{-1} \left( -\frac{\hat{\alpha}_{v,l}(t)}{m_r \omega_\text{nat} \sigma_{v,l}} \right) \quad \text{and} \quad
       \hat{\varphi}_{\text{env}}(t) = \kappa_u^{-1} \left( \frac{\hat{\alpha}_{u,l}(t)}{4m_r\omega_\text{nat}\sigma_{u,l}} \right) \quad \forall l \in \set{1,\dots,L}.
\end{align*}

Thus, by making the small-angle approximation in a post-processing step, we can use \cref{eq:expDecay} to set up the regression problem 
\begin{equation}
\hat{\varphi}_{\text{env}}(t) = \varphi_\text{max} \text{e}^{-\lambda t}.
\end{equation}
Taking the natural logarithm yields
\begin{equation}
\ln(\hat{\varphi}_{\text{env}}(t)) = \ln(\varphi_\text{max}\, \text{exp}\left(-\lambda t\right)) = \ln(\varphi_\text{max}) - \lambda t.
\end{equation}
Determining $\varphi_\text{max}$ and $\lambda$ from this equation is a linear regression problem \cite{freedman2009statistical} that can explicitly be solved by
\begin{align} \label{eq:linearRegressionDamping}
  \lambda^\text{SAE} & := -\frac{\int_0^T \left(\ln(\hat{\varphi}_{\text{env}}(t)) - b \right) \left(t - a \right) \text{d}t}{\int_0^T \left(t - a \right)^2 \text{d}t},  \qquad \varphi_\text{max}^\text{SAE} := \text{e}^{b + \lambda a},
\end{align}
where $a = \frac{T}{2}$ 
and $b = \underset{t \in [0,T]}{\text{mean}}(\ln(\hat{\varphi}_{\text{env}}(t)))$. 

\subsubsection{Optimization} \label{sec:optimization}
All methods considered in this work share the requirement to solve a nonlinear least-squares problem in order to estimate the parameters (see \cref{tab:overviewParam}). While the methods T$_\text{T}$ and TUSA$_\text{T}$ operate directly on the time-domain signals, the methods TUSA$_\text{F}$, TU$_\text{F}$, and TUSAE$_\text{F}$ are fitted to the instantaneous functions derived in the time-frequency domain. Since the optimization problems are multi-dimensional, nonlinear, and generally non-convex, convergence to the global minimum is not guaranteed; the cost function may contain local minima where the optimizer can become trapped. Therefore, careful selection of initial parameters is crucial to ensure reliable fitting.

Nonlinear least-squares problems can be approached with various optimization algorithms \cite{ruszczynski2011nonlinear}, broadly categorized as gradient-based or gradient-free methods. For problems of the type considered here, the Levenberg–Marquardt algorithm \cite{levenberg1944method} is a common choice, as it exploits the structure of the least-squares cost function to efficiently update parameter estimates. In each iteration, the Levenberg–Marquardt algorithm computes a gradient-based update that balances the rapid convergence of Gauss–Newton steps with the robustness of gradient descent. 

\subsubsection{Initial Parameters, Constraints and Weights}
\label{sec:startAndConstraints}

We consider different sets of initial parameters. We begin with a set of initial parameters $\mathcal{P}^\text{simp}$ using the time-frequency analysis (\cref{sec:timefreqanalysis}) and the model simplifications (\cref{subsec:TUSA}) to derive them directly from the instantaneous functions $\hat{\omega}(t)$, $\hat{\pmb{\alpha}}_v(t)$ and $\hat{\pmb{\alpha}}_u(t)$. Due to the relationship between the MC model and the TUSA model (see \cref{subsec:TUSA}), we directly obtain an estimator for the initial phase $\psi^\text{simp}_\text{start}$. As an initial estimate for the natural frequency, we use $\omega_\text{nat}^\text{simp} := \underset{t \in [0,T]}{\text{max}} \, \hat{\omega}(t)$.

The relaxation parameter $\lambda^\text{simp}$ is difficult to estimate from the time-frequency analysis, since the envelope function of the deflection angle $\varphi_\text{env}(t)$ is not available. Instead, we consider the amplitudes $|\hat{\alpha}_{v,l_\text{max}}(t)|$, where $l_\text{max}$ is the index of the channel with the strongest signal. We then determine $\lambda^\text{simp}$ by exponential curve fitting as in \cref{eq:linearRegressionDamping}. This heuristic approach is based on the observation that both the amplitudes and the deflection angle decrease at a similar rate, as they are strongly coupled. If we assume the function $\kappa_v$ to be linear in \cref{Eq:AvUndamped}, the amplitude would be proportional to $\varphi_\text{env}(t)$ and then the relaxation parameter would be identical.

An initial parameter for the maximum deflection angle $\varphi_\text{max}^\text{simp}$ can be obtained by calculating
\begin{align}
\varphi_\text{max}^\text{simp} := g^{-1}\left( \frac{\underset{t \in [0,T]}{\min}\hat{\omega}(t)}{\omega_\text{nat}^\text{simp}} \right),
\end{align}
i.e., we take the relative change in frequency and invert the function $g$ that connects $\varphi_\text{max}$ and $\omega$ in \cref{Eq:omegaUndampedFormula}.

Finally, based on the estimates of $\omega_\text{nat}^\text{simp}$ and $\varphi_\text{max}^\text{simp}$ and the instantaneous amplitudes 
$\hat{\pmb{\alpha}}_u(t)$ and $\hat{\pmb{\alpha}}_v(t)$,
the signal projections
$\pmb{\sigma}_u^\text{simp}$ and $\pmb{\sigma}_v^\text{simp}$ are derived  from \cref{Eq:AvUndamped} and \cref{Eq:AuUndamped} by computing
\begin{align}
    \pmb{\sigma}_v^\text{simp} 
    &:= - \frac{1}{m_\text{r}\,\kappa_v(\varphi_\text{max}^\text{simp})\,\omega_\text{nat}^\text{simp}}\,\hat{\pmb{\alpha}}_v(0) 
   \\[0.5em]
    \pmb{\sigma}_u^\text{simp}  
    &:= \frac{1}{4m_\text{r}\,\kappa_u(\varphi_\text{max}^\text{simp})\,\omega_\text{nat}^\text{simp}}\,\hat{\pmb{\alpha}}_u(0).
\end{align}
The initial parameters $\mathcal{P}^\text{simp}$
can be used for the methods in both the time domain and the time-frequency domain. For the time domain methods, specifically for method T$_\text{T}$, it is also possible to use the result of one of the time-frequency methods since the latter is very fast to calculate. In this work, we take the parameters from TUSA$_\text{F}$ to initialize the method T$_\text{T}$ and name these parameters $\mathcal{P}^\text{TUSA$_\text{F}$}$. Thus in the subsequent experimental section, we consider T$_\text{T}$$(\mathcal{P}^\text{simp})$ and T$_\text{T}$$(\mathcal{P}^\text{TUSA$_\text{F}$})$. An overview of the different methods, their estimation parameters, and corresponding initial parameters for the optimization is given in \cref{tab:overviewParam}.
\renewcommand{\arraystretch}{1.3}
\begin{table}[h!]
    \centering
    \begin{tabular}{lll}
    
    \textbf{Method} &Estimation Parameters &Initial Parameters\\
    \hline    
    T$_\text{T}$$(\mathcal{P}^\text{TUSA$_\text{F}$})$& $\{\pmb{\sigma}_u, \pmb{\sigma}_v, \omega_\text{nat}, \lambda, \varphi_\text{max}, \psi_\text{start} \}_\text{opt}$ & $\mathcal{P}^\text{TUSA$_\text{F}$}$  \\
    
     T$_\text{T}$$(\mathcal{P}^\text{simp})$& $\{\pmb{\sigma}_u, \pmb{\sigma}_v, \omega_\text{nat}, \lambda, \varphi_\text{max}, \psi_\text{start} \}_\text{opt}$ & $\mathcal{P}^\text{simp}$  \\

     TUSA$_\text{T}$ & $\{\pmb{\sigma}_u, \pmb{\sigma}_v, \omega_\text{nat}, \lambda, \varphi_\text{max}, \psi_\text{start} \}_\text{opt}$ & $\mathcal{P}^\text{simp}$   \\
     
       TUSA$_\text{F}$  & $\{\pmb{\sigma}_u, \pmb{\sigma}_v, \omega_\text{nat}, \lambda, \varphi_\text{max}\}_\text{opt} \cup \{\hat{\psi}_0\}_\text{STFT}$ &
       $\mathcal{P}^\text{simp}\setminus \{\varphi_\text{max}^\text{simp}$\}\\
       
        TUSAE$_\text{F}$  & $\{\pmb{\sigma}_u, \pmb{\sigma}_v, \omega_\text{nat}\}_\text{opt}\cup \{\hat{\psi}_0\}_\text{STFT} \cup \{\lambda^\text{SAE}, \varphi_\text{max}^\text{SAE}\}_\text{LR}$ & $\mathcal{P}^\text{simp}\setminus\{\omega_\text{nat}^\text{simp}, \lambda^\text{simp}, \varphi_\text{max}^\text{simp}\}$   \\
    \end{tabular}
    \caption{\textbf{Estimation and initial parameters for all methods.} The estimation parameters for  TUSA$_\text{F}$ are acquired by solving the optimization problem (opt) and from the STFT analysis. Using method TUSAE$_\text{F}$, an additional third estimation solving a linear regression problem (LR) is needed to get the complete parameter set $\mathcal{P}$.}
    \label{tab:overviewParam}
\end{table}
\renewcommand{\arraystretch}{1.0}

The constraints $\Gamma$ of the optimization problems are chosen as listed in \cref{tab:boxConstraints}. While the signal projections remain unconstrained, we force the phase $\psi_\text{start}$ to be in the interval $[0,2\pi)$. For the relaxation parameter $\lambda$, we enforce it to be positive and above $0.0005$, which is orders of magnitude lower than \acp{MMR} of typically size reach. The natural angular frequency $\omega_\text{nat}$ is constrained to not change by a factor of $4$ relative to the initial parameter $\omega_\text{nat}^\text{simp}$. Finally, the maximum deflection angle is constrained to be in the interval $[0,\SI{90}{\degree}]$. 

\renewcommand{\arraystretch}{1.4}
\begin{table}[h!]
    \centering
    \begin{tabular}{lcccccc}
    &$\psi_\text{start}$ &$\varphi_\text{max}$ &$\omega_\text{nat}$ &$\lambda$ &$\pmb{\sigma}_v$ &$\pmb{\sigma}_u$\\
    \hline
    Lower bound  &$0$       &\SI{0}{\degree}    &$0.25 \omega_\text{nat}^\text{simp}$      &$0.0005$ &$-\pmb{\infty}$&$-\pmb{\infty}$\\
    Upper bound  &$2\pi$        &\SI{90}{\degree}    &$4\omega_\text{nat}^\text{simp}$  &$\infty$ &$\pmb{\infty}$ &$\pmb{\infty}$\\
    \end{tabular}
    \caption{\textbf{Overview of the constraints} The constraints $\Gamma$ for the optimization problems of all methods, i.e., the lower and upper bounds for each parameter.}
    \label{tab:boxConstraints}
\end{table}
\renewcommand{\arraystretch}{1.0}
As method TUSA$_\text{F}$ optimizes the frequency and amplitude, we apply different weights to account for the different value ranges. The weights are calculated as follows: $\breve{w}(t) = 1/\text{Var}(\hat{\omega}(t))$ and $\breve{\pmb{W}}_v(t)$ = $\breve{\pmb{W}}_u(t) = 1/\text{Var}(\hat{\pmb{\alpha}}(t))$. All other methods use uniform weights for all data points. 

\subsection{Experiments}
\label{subsec:experimentalMethods}
\begin{figure}
\centering
    \includegraphics{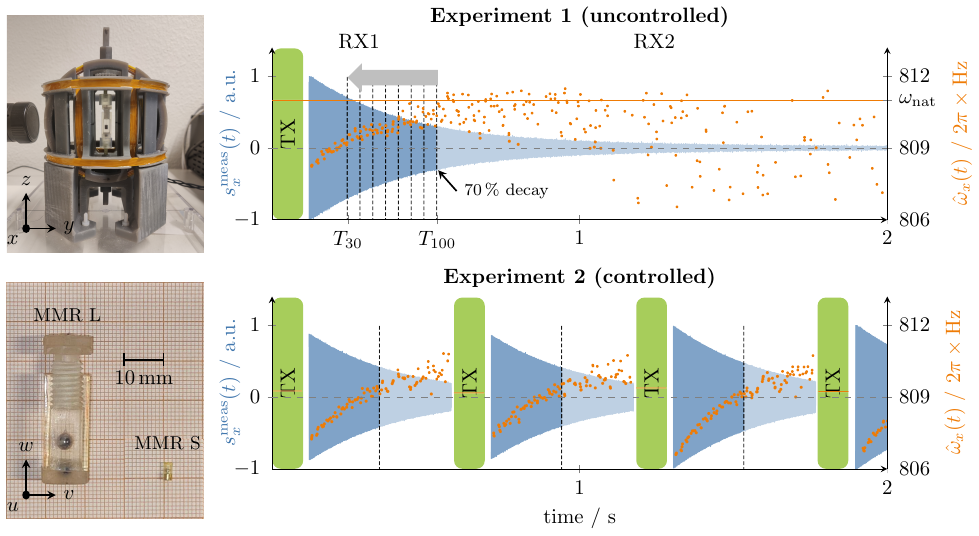}
    \caption{\textbf{Measurement setup, \acp{MMR}, and exemplary signals for the two experiment types (un-/controlled).} Left: 3D coil arrangement in Helmholtz configuration with the global coordinate system (top) and an image of \MMRL~and \MMRS~(bottom). Right: Exemplary filtered signals of \MMRS~measured with the $x$-coil. Orange dots indicate the instantaneous frequency acquired from the signal demodulation, the orange line during TX the excitation frequency of the current frame. The vertical dashed lines correspond to the cut-off times ranging from $T_\text{30}$ to $T_\text{100}$ for Experiment 1. The index denotes the proportion of the dataset considered, ranging from \SIrange{30}{100}{\percent}, which corresponds to the time interval over which the signal amplitude decreased by \SI{70}{\percent}. Note that in Experiment 1 the receive window was not divided into RX1 and RX2 during the measurement, however, in the analysis afterwards. For Experiment 2, a vertical dashed line indicates the RX1 and RX2 division for the control loop during measurement.}
    \label{fig:experimentalSetup}
\end{figure}

In order to assess the real-time capacity, accuracy, and precision of the methods, we performed experiments with our sensing system using two different \acp{MMR} (see \cref{fig:experimentalSetup}). 
Our sensing system consists of a 3D circular coil arrangement connected to an I/O card via a TX/RX switch.
The coils are arranged in Helmholtz configuration to ensure homogeneous excitation of the volume. They are used to generate the MMR excitation field and to measure the inductive sensor response. The global coordinate system ($x$,$y$,$z$) is defined as the coil axes. Each coil is made of thin litz wire with $50$ windings and different radii ($r_x = \SI{51.5}{\milli \meter}$, $r_y = \SI{60}{\milli \meter}$, and $r_z = \SI{68.5}{\milli \meter}$). The hardware setup for the receive and transmit path is similar to the one described by Faltinath et al.~\cite{faltinath2025natural}, with a minor adjustment concerning the TX/RX switch, which is presented by Mohn et al.~\cite{mohn2025transmit}. 

The larger \ac{MMR} (\MMRL) is composed of two spherical permanent magnets of neodymium-iron-boron and was already used by Faltinath et al.~\cite{faltinath2025natural}. The housing consists of 3D-printed components and is cuboidal with a mechanism to adapt the distance between the two magnets. The smaller \ac{MMR} (\MMRS) contains a spherical (rotor) and a cylindrical (stator) neodymium-iron-boron magnet. The housing is cylindrical and also 3D printed.  For both sensors a thin filament made of ultra-high-molecular-weight polyethylene is used to fix the rotor to the housing. Detailed specifications of the magnets used are listed in \cref{tab:settingsMagnets}.
\begin{table}[h]
    \centering
    \renewcommand{\arraystretch}{1.3}
    \begin{tabular}{lccc}
        \textbf{Parameter}      &\textbf{Spherical (\MMRL)}     &\textbf{Spherical (\MMRS)} &\textbf{Cylindrical (\MMRS)}\\
        \hline
         Size (diameter; height)    &\SI{4}{mm}; - &\SI{1}{mm}; -  &\SI{1}{mm}; \SI{1.5}{mm} \\
         Mass                       &\SI{350}{mg}  &\SI{31}{mg}   &\SI{35}{mg}\\             
         Remanence                  &\SIrange{1.26}{1.29}{\tesla} &\SIrange{1.42}{1.47}{\tesla} &\SIrange{1.40}{1.46}{\tesla}
    \end{tabular}
    \renewcommand{\arraystretch}{1.0}
    \caption{\textbf{Specifications of the manufacturer for the magnets used.} For \MMRL, the same type of magnet is used as both rotor and stator, whereas \MMRS~consists of a spherical rotor magnet and a cylindrical stator magnet.}
    \label{tab:settingsMagnets}
\end{table}

 Each sensor was centered within the coil setup and oriented such that the $v$-axis of the MMR was aligned with the $x$-axis and the $u$-axis with the $y$-axis. This was achieved by analyzing the signal in each coil with regard to the characteristics of the dedicated $v$-and $u$-signal. A measurement series (Experiment 1, uncontrolled) of $50$ frames was conducted. By employing a sinusoidal excitation signal in the $x$-coil, an external magnetic field was applied to achieve maximum torque at the equilibrium position. For all frames, the duration of the TX window was \SI{0.1}{s}. An additional short switch window of \SI{0.02}{s}, based on hardware constraints, was employed between the TX and RX window. The RX window duration was chosen to ensure that the signal decay transitioned into noise during the RX window, which resulted in durations of \SI{2}{s} for \MMRS~and \SI{10}{s} for \MMRL. The excitation amplitude was increased every 10 frames from \SIrange{15}{35}{\micro \tesla} in steps of \SI{5}{\micro \tesla} and the excitation frequency was fixed at \SI{811}{Hz} for \MMRS~and \SI{111}{Hz} for \MMRL. An overview of the measurement settings is shown in \cref{tab:overviewExperiments}.
 
\begin{table}[h]
\centering
\renewcommand{\arraystretch}{1.3}
\setlength{\tabcolsep}{5pt}
\begin{tabular}{ccccccccc}
 Experiment  &controlled &MMR &TX\,/\,\si{s}  &RX\,/\,\si{s}  &$\omega_\text{TX}$\,/\,$2\pi \times \si{Hz}$ &$B_\text{TX}$\,/\,\si{\micro\tesla} &cut-off times & $\varphi_\text{max}$ \\
 \hline
\multirow{2}{*}{1} &no &S   &0.10 &2.0   &811   &\SIrange{15}{35}{}           &$T_\text{30}$ to $T_\text{100}$ &\SIrange{7.3}{10.9}{\degree}             \\
    &no& L   & 0.10 & 10.0  & 111   &\SIrange{15}{35}{}  & $T_\text{30}$ to $T_\text{100}$ & \SIrange{7.6}{13.5}{\degree}  \\
   \multirow{2}{*}{2}   &yes& S   & 0.07 & 0.3 & dyn.  & \SIrange{35}{155}{}          & $T_\text{100}$  &\SIrange{7.6}{28.8}{\degree}              \\
   &yes& L   & 0.10 & 2.0 & dyn.              &\SIrange{15}{100}{}    & $T_\text{100}$  &\SIrange{8.8}{29.4}{\degree}
\end{tabular}
\renewcommand{\arraystretch}{1.0}
\caption{\textbf{Overview of the measurement settings.} The different settings of both experiments, as well as the maximum deflection angle.}
\label{tab:overviewExperiments}
\end{table}
 With the aforementioned measurement series (Experiment 1, uncontrolled), only maximum deflection angles of $\le \SI{14}{\degree}$ are achieved as the movement of the rotor decayed fully in each frame. To analyze the deviation for the simplified models on a measurement with larger deflection angles, we conducted experiments with both \acp{MMR} where the rotor was increasingly re-excited (Experiment 2, controlled). Thus, the small-angle approximation is no longer valid. The TX window was adapted for \MMRS~and the RX window shortened to a time when the rotor still oscillates. The excitation amplitude was again increased every $10$ frames in steps of \SI{5}{\micro \tesla} (see \cref{tab:overviewExperiments}). In the first frame, the phase shift was zero and the frequency set to \SI{811}{Hz} for \MMRS~and \SI{111}{Hz} for \MMRL. In the following frames, a parameter estimation during the RX2 window was conducted with the method TUSA$_\text{F}$ to acquire $\psi_\text{TX}(T)$ and $\omega_\text{TX}(T)$ for the excitation in the next TX window.

Before estimating the parameters with the different methods, the measured data of both experiments were first post-processed with a digital fourth-order Butterworth bandpass filter to suppress noise. The passband encompasses the \ac{MMR}'s natural frequency and its second harmonic. For \MMRL, the passband ranged from \SI{80}{Hz} to \SI{300}{Hz}, and for \MMRS, from \SI{600}{Hz} to \SI{2000}{Hz}. For each frame, several STFTs with Gaussian windows of different lengths ($T_\text{snip}$) were performed, depending on the period of one MMR oscillation ($T_\text{MMRS}$ or $T_\text{MMRL}$). For \MMRS, it was $T_\text{snip} = T_\text{step} = \{14,15,16,17\} \cdot T_\text{MMRS}$, and for \MMRL, $T_\text{snip} = T_\text{step} = \{9,10,11\} \cdot T_\text{MMRL}$, as those seemed to produce the least fluctuations on the frequency trace while being as short as possible to fulfill the assumption that the amplitude and frequency change are small within each bin of the STFT. In addition, a \ac{SNR} threshold of $1.5$ for \MMRS~and $1.0$ for \MMRL~was applied to $\hat{\omega}, \hat{\pmb{\alpha}}_v, \hat{\pmb{\alpha}}_u$. The voltage signal and instantaneous frequencies acquired from the STFT of example frames can be seen in \cref{fig:experimentalSetup} on the right. 

To analyze the accuracy of the parameter estimation for shorter RX1 windows, truncated portions of the data from Experiment 1, ranging from \SIrange{30}{100}{\percent} (data30 to data100), were used to estimate the parameter set $\mathcal{P}$. The time $T_\text{100}$ (data100) was set to an amplitude decay of \SI{70}{\percent} of each frame, separating RX1 and RX2  in the analysis. The naming of the datasets is as follows: dataX$_\text{Y}$, where X denotes the used percentage ranging from \SIrange{30}{100}{\percent} and Y denotes whether the smaller (S) or the larger (L) MMR was used. The cut-off times $T_\text{30}$ to $T_\text{100}$ are \SIrange{0.13}{0.49}{s} for \MMRS~and \SIrange{0.61}{2.66}{s} for \MMRL~and are exemplarily illustrated in \cref{fig:experimentalSetup}. As can be seen, the instantaneous frequencies scatter significantly after $T_\text{100}$.

Regarding the control parameters, we focus in this study on the frequency $\omega_\text{TX} = \omega^\text{model}(T)$ and phase $\psi_\text{TX} = \psi^\text{model}(T)$ of the subsequent frame as those have to be forecast. Since method T$_\text{T}$ does not directly provide an instantaneous frequency and phase, we employ the STFT method discussed in \cref{sec:timefreqanalysis}. Subsequently, we derive $ \omega^\text{T}(T)$ and $\psi^\text{T}(T)$.

All methods were implemented in the programming language Julia (version 1.10). To solve the \ac{ODE} for method T$_\text{T}$ the explicit Runge-Kutta method Tsit5 of the package \texttt{OrdinaryDiffEq.jl}\footnote{\scriptsize{\url{https://github.com/SciML/OrdinaryDiffEq.jl}}} (version 6.93.0) was used. The absolute and relative error tolerances were both set to $10^{-8}$. The nonlinear least-squares problem was solved using the package \texttt{LsqFit.jl}\footnote{\scriptsize{\url{https://github.com/JuliaNLSolvers/LsqFit.jl}}} (version 0.15.0), which allows box constraints to be applied during optimization. The gradients are approximated using forward finite differences, the package's default tolerances taken and the optimization is terminated after 100 iterations.

The runtime of each dataset (data30$_\text{S/L}$ - data100$_\text{S/L}$) was measured by evaluating each method five times on every frame. Afterwards, the mean and standard deviation over all frames were computed, yielding the final runtime statistics for each method and dataset. The software package \texttt{BenchmarkTools.jl}\footnote{\scriptsize{\url{https://github.com/JuliaCI/BenchmarkTools.jl}}} (version 1.6.0) was used to measure the runtime on a workstation with an AMD Ryzen 7 PRO 7730U, a processor base frequency of \SI{2}{\giga \hertz}, and \SI{64}{\giga \byte} main memory.

\subsection{Evaluation Metrics} \label{sec:evaluationmetrics}
To assess the quality of the reconstructed signals using the different methods, we calculated the \ac{NRMSE} between the measured signal and the reconstructed signal, normalized to the mean value of the measured signal in each channel. 
As the torsion model T$_\text{T}$$(\mathcal{P}^\text{TUSA$_\text{F}$})$ initialized with the parameters from method TUSA$_\text{F}$ using data100 has the least simplifications, we declare it as our reference method.

To evaluate the performance of the methods, either for truncated RX1 windows or larger maximum deflection angles, the estimates for the parameter set $\mathcal{P}$ were compared to those of the reference method. 
Relative deviations were computed for the deflection angle, natural frequency, and relaxation parameter. The phase deviation was quantified by the absolute deviation. For the signal projections, we analyzed the alignment as well as the magnitude. The former was evaluated using the cosine similarity $\text{sim}_\text{cos}$ \cite{deza2016errorMetrics} and the latter by calculating the relative deviation. 

Concerning the control parameters, the frequency and phase of the subsequent frame have to be forecast. We therefore analyze the deviation to the reference method for one frame. For the frequency the relative deviation was calculated whereas the phase deviation is quantified using the Euclidean distance D$_\text{C} (\psi(t))$, where $0$ indicates no deviation and $2$ represents a \SI{180}{\degree} phase shift, to avoid phase wraps.

We further analyzed the reproducibility of the sensing parameters ($\omeganat, \pmb{\sigma}_v, \pmb{\sigma}_u$) across several frames on Experiment 1 - data100 by calculating the  mean and corrected sample standard deviation for each method and sensing parameter. Since position, orientation, and natural frequency remained constant during the MMR measurements presented in \cref{subsec:experimentalMethods}, consistent estimates are expected. 

\section{Results}
\label{sec:results}
\begin{figure}[t!]
    \centering
    \includegraphics{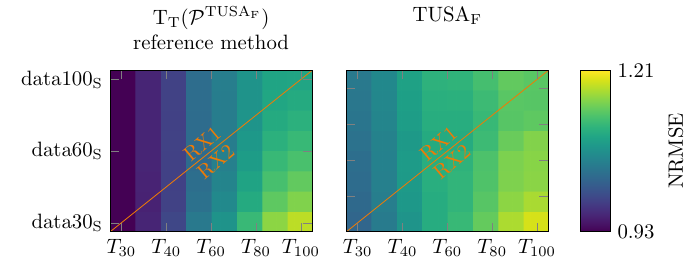}
    \caption{\textbf{Error to time signal; Experiment 1 - uncontrolled (\MMRS).} The normalized root mean square error for the reference method (left) and method \text{TUSA$_\text{F}$} (right) over all frames are shown for different amounts of data and for different time ranges.}
    \label{fig:errorHeatmapTimeSignal}
\end{figure}
In the following, we mainly focus on the results for \MMRS, as they are representative for \MMRL~as well. We first show the error between the measured signal and the reconstructed signal for both the reference method and the method applied in Experiment 2 - controlled. We then evaluate the methods with respect to their suitability concerning controlling and real-time estimation. Subsequently, the precision and uncertainty for the sensing parameters ($\omeganat, \pmb{\sigma}_v, \pmb{\sigma}_u$) are presented. Finally, the performance for larger deflection angles is displayed.

\cref{fig:errorHeatmapTimeSignal} shows the \acp{NRMSE} between the measured and reconstructed signals for two methods. Method TUSA$_\text{F}$ exhibits a higher error than method T$_\text{T}$$(\mathcal{P}^\text{TUSA$_\text{F}$})$. Both methods show reduced accuracy when reconstructing the measured signal later in RX2. The \acp{NRMSE} for the other methods are shown in \cref{sup_fig:errorHeatmapTimeSignal_MMRS}, however, they exhibit similar trends. The methods working in the time domain produce similar error matrices and method TUSAE$_\text{F}$ has the highest error. The \ac{NRMSE} for estimating the measured signal \MMRL~is generally lower, yet shows a similar trend (see \cref{sup_fig:errorHeatmapTimeSignal_MMRL}).

\begin{figure}[t!]
    \centering
    \includegraphics{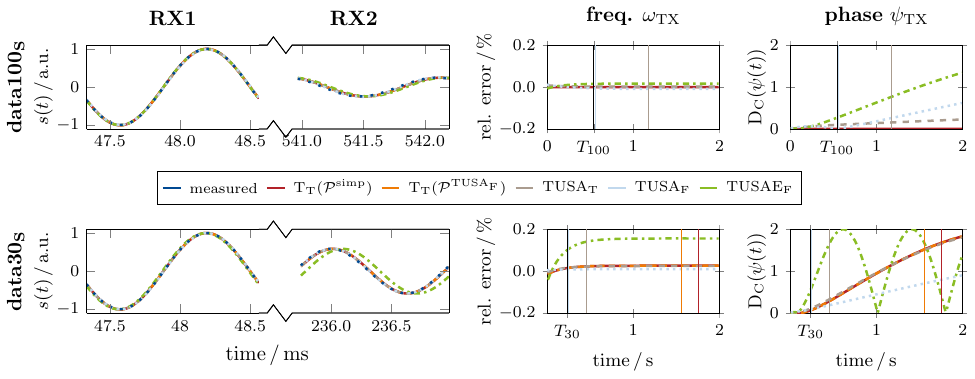}
    \caption{\textbf{Forecasts of time signal and control parameter; Experiment 1 - uncontrolled (\MMRS).} Left: Exemplary signal section at different time points of the measured signal in comparison to the estimated signals of all methods (excitation amplitude: \SI{35}{\micro T}). Signals for a time segment during RX1 as well as $\SI{100}{\milli s}$ after $T_\text{30}/T_\text{100}$ (labeled RX2) are shown. Note that, all methods reconstruct the signal very closely to the measured signal within RX1. Right: The deviation for all methods from the reference method for data30$_\text{S}$ and data100$_\text{S}$ for the forecast of frequency $\omega_\text{TX}$ and phase $\psi_\text{TX}$ over time is displayed. Additionally, the cut-off time $T_\text{30}$, $T_\text{100}$ as well as the mean runtime of the methods are shown in solid lines. As the mean runtime of the methods T$_\text{T}$ for data100$_\text{S}$ are longer than the measured RX window, they are not displayed.}
    \label{fig:fittedSignals_controlParams}
\end{figure}

In \cref{fig:fittedSignals_controlParams} on the left, exemplary signal sections for the estimated and post-processed measured data at different periods of time are shown. All reconstructed signals are close to the measured signal when data100$_\text{S}$ was used to estimate the parameters. When data30$_\text{S}$ was used, the reconstructed data is similar during RX1. However, looking at the reconstructed signals for a time period which is shortly after RX1, the signal from method TUSAE$_\text{F}$ is already defective with regard to the phase. 
For the control signal of the subsequent frame, the methods are required to predict both frequency $\omega_\text{TX}$ and phase $\psi_\text{TX}$. In \cref{fig:fittedSignals_controlParams} on the right, deviations for all methods from the reference method for data30$_\text{S}$ and data100$_\text{S}$ are displayed for one RX window. The forecasts were performed over the entire frame. As can be observed, the time-domain methods T$_\text{T}$ do not complete their estimation within the \SI{2}{s}. In contrast, the time-frequency-based methods are almost instantly finished, allowing for shorter RX2 windows. 
Consequently, they have lower or similar deviation (except for method TUSAE$_\text{F}$ for the frequencies) than the time-based methods at the end of their respective RX2 window. With respect to \MMRL~it is noteworthy that, method TUSAE$_\text{F}$ performs comparably to the other methods for the frame shown (see \cref{sup_fig:fittedSignals_controlParams_MMRL}).

\begin{figure}[t]
    \centering
    \includegraphics{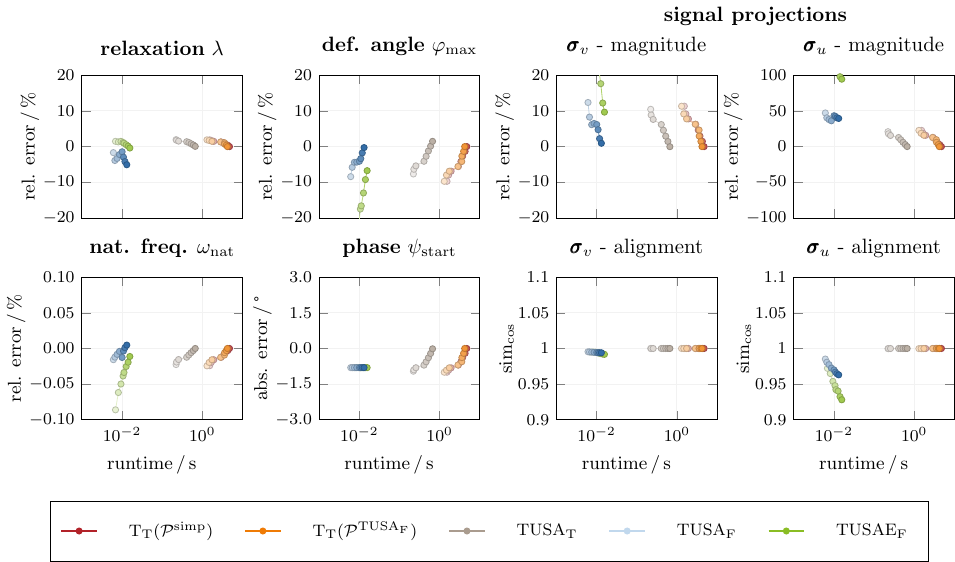}
    \caption{\textbf{Runtime and deviation for different datasets; Experiment 1 - uncontrolled (\MMRS).} The median deviation between the different methods and the reference method of all frames over the runtime for the truncated receive windows. For the parameters $\lambda,\varphi_\text{max}, \omeganat, \pmb{\sigma}_{v}$ and $\pmb{\sigma}_{u}$, the deviation is quantified using the relative error, for $\psi_\text{start}$ using the absolute error, and for the alignment of $\pmb{\sigma}_{v}$ and $\pmb{\sigma}_{u}$ using the cosine similarity. The color fading in the line plots indicates the amount of data used, with the lightest shade representing data30$_\text{S}$ and the darkest data100$_\text{S}$.}
    \label{fig:MMRS_Experiment1}
\end{figure}

To comparatively evaluate the performance of the different methods on truncated receive windows, we evaluate the needed runtime and median deviation of the estimated parameters relative to T$_\text{T}$($\mathcal{P}^{\text{TUSA}_\text{F}}$) with data100$_\text{S}$ (see \cref{fig:MMRS_Experiment1}). 
In general, the estimations done in the time domain are slower with T$_\text{T}(\mathcal{P}^\text{simp})$ being the slowest, ranging from $t_\text{mean} =\SI{5.0}{s}$ for data100$_\text{S}$ to \SI{1.5}{s} for data30$_\text{S}$. A slight improvement of around \SI{0.2}{s} is achieved by using a different set of initial parameters $\mathcal{P}^{\text{TUSA}_\text{F}}$. The runtime of method TUSA$_\text{T}$ has decreased by around sixfold. The methods operating in the time-frequency domain are the most efficient, being a factor of $40$ faster than TUSA$_\text{T}$. For \MMRL~a similar tendency was observed in the runtime, however with a generally higher runtime. In addition, with data40$_\text{L}$ and lower, the mean runtime of the time-based methods increased again~(see \cref{sup_fig:MMRL_Experiment1}).

Concerning the deviations between the estimated parameters, it increased most of the time when less data was used, with the time methods exhibiting a similar trend for all parameters. The maximum deflection angle is usually underestimated while the magnitudes of the signal projections are usually overestimated when using less data. Those are also the parameters where the deviation increases the most, especially for method TUSAE$_\text{F}$, exceeding all others with data50$_\text{S}$ and less. The phase, natural frequency, and the alignment of the signal projection vectors have the lowest deviation.

Again, a similar trend was seen for the results with \MMRL~with the biggest difference being that all methods performed worse concerning the signal projection $\pmb{\sigma}_{u}$. In addition, the relative deviation for the relaxation parameter of method TUSA$_\text{F}$ was high for all data sets.

In \cref{tab:sensingResults}, the mean and standard deviation of the methods for the sensing parameters for both MMRs are shown. The natural frequency has the same precision for all methods for the entire measurement. However, a closer inspection of the estimated natural frequency reveals a decreasing trend with increasing excitation amplitude~(see \cref{sup_fig:fnatShift}). For both \acp{MMR}, the estimated natural frequency of the methods in the time domain decreases when the excitation amplitude increases. The methods in the time-frequency domain exhibit a similar overall trend, however, with more outliers.

As both MMRs were oriented such that the $v$-axis mainly pointed in $x$-axis and $u$-axis mainly in $y$-axis, the highest absolute values are assumed in the corresponding pairs. This is also visible in the mean values of all methods for $\pmb{\sigma}_v$, with the frequency methods having slightly different values and lower precision. For $\pmb{\sigma}_u$, the time domain method also estimated the corresponding pairs, however, there is a high value in the $z$-direction for \MMRS~but not for \MMRL. The methods operating in the time-frequency domain had an overall higher uncertainty compared to the methods operating in the time domain in estimating $\pmb{\sigma}_u$.

\begin{figure}[h]
    \centering
    \includegraphics{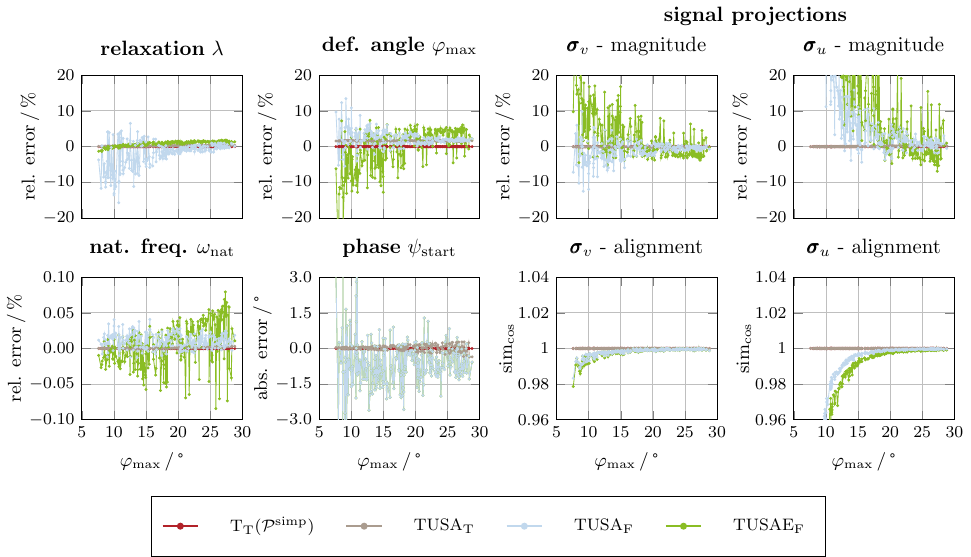}
    \caption{\textbf{Deviation for larger deflection angles; Experiment 2 - controlled (\MMRS).} The deviation of the different methods from the reference method (T$_\text{T}$$(\mathcal{P}^\text{TUSA$_\text{F}$})$) over the maximum deflection angle. For the parameters $\lambda,\varphi_\text{max}, \omeganat, \pmb{\sigma}_{v}$ and $\pmb{\sigma}_{u}$ the deviation is quantified using the relative error, for $\psi_\text{start}$ using the absolute error, and for the alignment of $\pmb{\sigma}_{v}$ and $\pmb{\sigma}_{u}$ using the cosine similarity.}
    \label{fig:MMRS_Experiment2}
\end{figure}

\begin{table}[h]
\renewcommand{\arraystretch}{1.3}
\setlength{\tabcolsep}{5pt}
    \centering
    \begin{tabular}{clccccccc}
    \hline
    &\textbf{Method} &\textbf{nat.\ freq.\,/\,\si{\hertz}} &\multicolumn{3}{c}{\textbf{signal projection} $\pmb{\sigma}_v$\,/\,\si{\tesla \per \ampere}} &\multicolumn{3}{c}{\textbf{signal projection} $\pmb{\sigma}_u$\,/\,\si{\tesla \per \ampere}}\\
      & & &$x$ &$y$&$z$&$x$&$y$&$z$\\
      \hline
\multirow{5}{*}{\rotatebox{90}{\textbf{\MMRS}}}&T$_\text{T}$$(\mathcal{P}^\text{simp})$ &$810.6\pm0.2$                 &$250\pm30$	    &$6.5\pm0.8$	&$2.0\pm0.4$	&$3\pm2$	&$-160\pm40$	&$-120\pm20$ \\
&T$_\text{T}$$(\mathcal{P}^\text{TUSA$_\text{F}$})$ &$810.6\pm0.2$                 &$250\pm30$	    &$6.5\pm0.8$	&$2.0\pm0.4$	&$3\pm2$	&$-160\pm40$	&$-120\pm20$ \\
&TUSA$_\text{T}$ &$810.6\pm0.2$        &$240\pm30$	    &$6.5\pm0.8$	&$2.0\pm0.4$	&$3\pm2$	&$-160\pm40$	&$-120\pm20$ \\
&TUSA$_\text{F}$ &$810.6\pm0.2$   &$250\pm30$	&$-1\pm4$	&$-30\pm20$	&$-1\pm5$	&$-180\pm40$	&$-200\pm300$ \\
&TUSAE$_\text{F}$ &$810.5\pm0.2$  &$280\pm30$	&$-1\pm4$	&$-30\pm30$	&$-1\pm5$	&$-210\pm50$	&$-400\pm600$ \\
\hline
\multirow{5}{*}{\rotatebox{90}{\textbf{\MMRL}}}&T$_\text{T}$$(\mathcal{P}^\text{simp})$   &$113.35\pm0.03$	&$190\pm40$	    &$43\pm9$	    &$1.9\pm0.5$	&$21\pm9$	&$-70\pm30$	    &$6\pm3$\\
&T$_\text{T}$$(\mathcal{P}^\text{TUSA$_\text{F}$})$  &$113.35\pm0.03$	&$190\pm40$	    &$43\pm9$	    &$1.9\pm0.5$	&$21\pm9$	&$-70\pm30$	    &$6\pm3$\\
&TUSA$_\text{T}$   &$113.35\pm0.03$	&$190\pm40$	    &$43\pm9$	    &$1.9\pm0.4$ &$20\pm9$     &$-70\pm30$	    &$6\pm3$\\
&TUSA$_\text{F}$             &$113.37\pm0.03$	&$180\pm40$     &$41\pm8$	    &$-10\pm20$	&$-1\pm4$	   &$-110\pm70$	    &$-100\pm200$\\
&TUSAE$_\text{F}$            &$113.36\pm0.03$	&$200\pm40$     &$44\pm8$	    &$-10\pm20$	&$-1\pm5$	   &$-200\pm200$	    &$-200\pm200$\\

\hline
    \end{tabular}
    \caption{\textbf{Precision of the sensing parameters; Experiment 1 - uncontrolled (data100).} Mean and corrected sample standard deviation over all frames for all methods for the sensing parameters (natural frequency and signal projections).}
    \label{tab:sensingResults}
    \renewcommand{\arraystretch}{1}
\end{table}
The performance of the simplified methods for increasing deflection angles is visualized in \cref{fig:MMRS_Experiment2}. Here, the parameter deviation from the estimation with T$_\text{T}$$(\mathcal{P}^\text{TUSA$_\text{F}$})$ is shown over the maximum deflection angles (estimated with T$_\text{T}$$(\mathcal{P}^\text{TUSA$_\text{F}$})$). Except for the parameter $\psi_\text{start}$ and for method TUSAE$_\text{F}$, the deviation and, consequently, the estimated parameters have fewer outliers for larger $\varphi_\text{max}$. Also, a large reduction of the relative error for $\lambda$ estimated with method TUSA$_\text{F}$ can be seen. Similar trends can be seen in \cref{sup_fig:MMRL_Experiment2} for \MMRL. There are two notable differences: fewer fluctuations in the deviations of the time-frequency domain methods and the deviation for $\varphi_\text{max}$ grows with increasing $\varphi_\text{max}$ for the simplified methods.

\section{Discussion}
We presented several methods relevant for controlling and sensing with magneto-mechanical resonators and evaluated their performance concerning temporal resolution, accuracy, and precision for different maximum deflection angles. 
As there is no real ground truth available, the torsion model in the time domain (T$_\text{T}$) serves as a reference method as it has the least simplifications. However, it relies on the assumption that the rotor only performs a 1D torsional motion. In practice, additional parasitic modes, such as pendulum-like oscillations, may be present. Future work should therefore examine parasitic modes in more detail and assess whether they can be incorporated into an extended model for parameter estimation. 

The main advantage of all simplified models is their computational efficiency: unlike the reference method, they do not require solving a differential equation, but only rely on explicit function evaluations. This efficiency comes at the cost of additional assumptions; however, the results indicate that they can be used to estimate the parameter set $\mathcal{P}$. In the case of the undamped small-angle model in the time domain (TUSA$_\text{T}$), the deviation from the reference method is $\le$ \SI{2}{\percent}. For maximum deflection angles of $\ge$ \SI{20}{\degree}, the undamped small-angle model in the time-frequency domain (TUSA$_\text{F}$) achieved $\le$ \SI{5}{\percent} deviation and the undamped model with small-angle extension in time-frequency domain (TUSAE$_\text{F}$) achieved $\le$ \SI{20}{\percent}. The high value is mostly due the deviation of the signal projection of $\pmb{\sigma}_u$.

Concerning the results in terms of runtime and accuracy while meeting the requirements of the small-angle approximation, method TUSA$_\text{T}$ performed best. Using it on data80 instead of the complex method T$_\text{T}$ on data100, it is possible to reduce the RX window from \SI{5.5}{s} to \SI{1.0}{s} with a deviation of less than \SI{5}{\percent} for all parameters. Moving the optimization problem into the time-frequency domain leads to a reduction of data inherent to the STFT representation and consequently to a further runtime reduction. On the one hand, this increases the deviation of the parameters from the reference method. On the other hand, the significantly faster runtime allows for shorter RX2 windows. Consequently, their short-time accurate frequency and phase estimates are sufficient for the subsequent control signal. However, reducing the data points in the time signal may also improve the runtime of the methods working in the time domain. A comprehensive analysis of the impact of reducing the data points in the time signal on the runtime and performance must be addressed in a future study.

In addition to frequency and phase, the amplitude of each coil would also be needed for a sinusoidal control signal. In the experiments conducted in this study, the MMR remained static and positioned within a homogeneous magnetic field. However, if the MMR moves freely within the field-of-view of a planar inductive coil array, as illustrated in \cref{fig:MMRBasics}, it would be necessary to adjust the amplitude of each coil accordingly. They could be determined using the signal projection and the maximum deflection angle. However, this is a control problem that will be addressed in future work.

Concerning the parameters needed for sensing an environmental parameter, i.e. the natural frequency, it can be seen that the deviation from the reference estimation is less than \SI{0.1}{\percent}. The precision of all methods for \MMRS~is \SI{0.2}{Hz} and for \MMRL~even less (\SI{0.03}{Hz}), indicating the lower limit for the sensitivity of a dedicated sensing application and MMR. The precision appears to depend on the MMR, which may be related to an observed drift in the estimated sensor frequency. The estimated sensor frequency decreased slowly as the maximum deflection angle increased. Therefore, the effect of the maximum deflection angle needs to be analyzed separately.

The estimation of the signal projections performed worse, especially in the magnitude, leading to the conclusion that estimating the position and orientation can lead to deviations from the true MMR pose. Mostly, the methods operating in the time-frequency domain cannot be used for tracking in the current state while fulfilling the small-angle approximation. The highest uncertainty was observed in the $z$-direction. This may be due to the fact that among all coils, the one measuring in $z$-direction had the lowest SNR. The scope of the study did not include solving the inverse pose estimation problem, therefore, the actual impact on the estimated pose cannot be quantified. This should be investigated in a subsequent, dedicated tracking study involving larger maximum deflection angles ($> \SI{20}{\degree}$). 

As the simplified models assume the small-angle approximation, the time signal and the parameters describing it are expected to deviate from the exact dynamics for large deflection angles. Despite this, the deviation, especially for the methods in the time-frequency domain, becomes less with increasing maximum deflection angle. The parameter deviation of method TUSA$_\text{F}$ decreased the most for larger $\varphi_\text{max}$. The current limitation of those methods therefore seems to be a low \ac{SNR} of the signal for smaller deflection angles, rather than the small-angle approximation. 

In addition to the parameter-estimation method, the repetition rate for sensing further depends on the relaxation parameter of the MMR. In this study, \MMRS~had a higher relaxation parameter, causing a faster decrease of the amplitude compared to \MMRL. Consequently, it must be re-excited more frequently to keep the MMR in an excited state. During the TX window, no signal from the MMR can be measured with the current setup. It is therefore favorable to have a short TX window and/or excite less frequently. If the MMR has a low relaxation parameter, the rotor remains longer in an oscillatory state. Provided this duration is sufficient, the RX window could be subdivided into multiple segments, each with its own parameter estimation. This could result in several estimations of the sensing parameters per frame. However, this concept was not addressed in this study and will be considered in future work.

When using an STFT, there is always a trade-off between temporal resolution and frequency resolution. In addition, it has to be ensured that the amplitude and frequency change is small during an STFT window. Consequently, as the relaxation parameter was lower for \MMRL, the STFT window can be longer than for \MMRS. We determined the used window length heuristically. While the results are useful, the approach is still preliminary and could be enhanced by defining criteria depending on the quality factor, the \ac{MMR}'s instantaneous frequency, and the SNR. As the STFT windows were short, we used a Gaussian spectrum interpolation to improve the instantaneous frequency estimate. Using more than three points for the Gaussian interpolation, constraining the instantaneous frequency of the next STFT window based on the previous one, or using other methods like, e.g., the wavelet transform, could further improve the time-frequency estimation methods.

Since the primary aim of this paper was to compare the different methods and illustrate general trends in runtime and accuracy, the implemented algorithms were not fully optimized for dedicated scenarios. For example, using prior knowledge could be applicable to reduce the dimensionality of the optimization problem. If certain parameters are static, they can be estimated offline in advance. For example, this could be the case in a dedicated tracking experiment that does not involve measuring an environmental parameter via the natural frequency. 

\section{Conclusion}
We have presented several models to describe the measured signal as well as methods to estimate the parameters for those models. The simplified methods have great potential to improve the temporal resolution of a measurement using magneto-mechanical resonators. They capture all essential features of the measured signal: the instantaneous frequency increases over time, while the maximum deflection angle and thus the signal amplitudes decays exponentially. Although time-domain methods are more accurate, they are computationally more complex. In contrast, the time-frequency domain methods are less accurate but reduce the temporal duration of the control forecast window by two orders of magnitude, leading to reliable results of the two control parameters frequency and and phase, which have to be predicted for the subsequent frame. The simplified methods can therefore contribute towards controlling and sensing with magneto-mechanical resonators in real-time. For now, we recommend to use the undamped small-angle model in the time-frequency domain for controlling and sensing an environmental parameter and the undamped small-angle model in the time domain in combination with a dedicated tracking algorithm for sensing the MMR's pose. 

\section*{Data availability}
The experimental data that support the findings of this study are openly available at \url{https://doi.org/10.15480/882.16742} \cite{reiss2026dataSet}. 

\section*{Code availability}
The implementation of the torsional model was written in Julia and is published under the MIT License at \url{https://github.com/IBIResearch/MMR-Parameter-Estimation}.
\section*{Acknowledgments}
This project is funded by the Deutsche Forschungsgemeinschaft (DFG, German Research Foundation) - SFB 1615 - 503850735.

\section*{Competing interests}
All other authors declare no competing interesting.

\printbibliography[]

\appendix
\renewcommand{\thefigure}{\Alph{section}.\arabic{figure}}
\setcounter{figure}{0}

\section{Appendix}
\label{app:STFTSpectralRepresentation}
In this appendix, the explicit expression \cref{eq:STFTCompInFS} of the STFT for the signal $s_{j,n}(t) = \frac{1}{2}\alpha_{j,n}  \text{e}^{\mathrm{i}( \omega_{j,n} t + \psi_{j,n} )}$ is derived. Inserting the signal into the STFT definition \cref{eq:STFT} yields
\begin{align*}
     S_w^{j,n}(t, \tilde{\omega}) &=  \int _{-\infty }^{\infty } s_{j,n}(\tau) w(\tau-t )\mathrm {e} ^{-\mathrm{i} \tilde{\omega} \tau}\,\mathrm {d} \tau \\
     &= \frac{\alpha_{j,n}}{2} \text{e}^{\mathrm{i} \psi_{j,n}} \int _{-\infty }^{\infty }   \text{e}^{\mathrm{i} \omega_{j,n} \tau } w(\tau - t )\mathrm {e} ^{-\mathrm{i} \tilde{\omega} \tau}\,\mathrm {d} \tau.
\end{align*}
Substituting $\tau' := \tau - t$ leads to 
\begin{align*}     
     S_w^{j,n}(t, \tilde{\omega}) & = \frac{\alpha_{j,n}}{2} \text{e}^{\mathrm{i} \psi_{j,n}} \int _{-\infty }^{\infty }   \text{e}^{\mathrm{i} \omega_{j,n} (\tau'+t) } w(\tau' )\mathrm {e} ^{-\mathrm{i} \tilde{\omega} (\tau'+t)}\,\mathrm {d} \tau' \\
      & = \frac{\alpha_{j,n}}{2} \text{e}^{\mathrm{i} ((\omega_{j,n} - \tilde{\omega}) t + \psi_{j,n}  )} \int _{-\infty }^{\infty }   \text{e}^{\mathrm{i} \omega_{j,n} \tau' } w(\tau' )\mathrm {e} ^{-\mathrm{i} \tilde{\omega} \tau'}\,\mathrm {d} \tau' \\
      &= \frac{\alpha_{j,n}}{2} \text{e}^{\mathrm{i} ((\omega_{j,n} - \tilde{\omega}) t + \psi_{j,n}  )} {\cal F} \{ \text{e}^{\mathrm{i} \omega_{j,n}  \,\cdot\,} w(  \,\cdot\, ) \}(\tilde{\omega}).
\end{align*}
Here, we consider the non-unitary, angular frequency Fourier transform ${\cal F} \{ a \}(\tilde{\omega}) := \int _{-\infty }^{\infty }   a(t) \mathrm {e} ^{-\mathrm{i} \tilde{\omega} t}\,\mathrm {d}t$ with its inverse ${\cal F}^{-1} \{ {\cal F} \{ a \} \}(t) := \frac{1}{2\pi}\int _{-\infty }^{\infty }  {\cal F} \{ a \} \mathrm {e} ^{\mathrm{i} \tilde{\omega} t}\,\mathrm {d}\tilde{\omega}$. Using the convolution theorem ${\cal F} \{ a \, b \}(\tilde{\omega}) = \frac{1}{2\pi} {\cal F} \{ a \} \ast {\cal F} \{ b \}(\tilde{\omega})$ and with ${\cal F} \{ \text{e}^{\mathrm{i} \omega_{j,n}  \,\cdot\,} \}(\tilde{\omega}) = 2\pi \delta(\tilde{\omega} - \omega_{j,n}) $, where $\delta$ is the Dirac delta distribution, and ${\cal F} \{w\}(\tilde{\omega}) = W(\tilde{\omega})$, we obtain
\begin{align*}     
     S_w^{j,n}(t, \tilde{\omega}) & = \frac{\alpha_{j,n}}{2} \text{e}^{\mathrm{i} ((\omega_{j,n} - \tilde{\omega}) t +\psi_{j,n}  )} 
     \frac{1}{2\pi}\int_{-\infty}^{\infty}  2\pi \delta(\omega' - \omega_{j,n}) W(\tilde{\omega} - \omega')     \text{d}\omega' \\
     &= \frac{\alpha_{j,n}}{2} \text{e}^{\mathrm{i} ((\omega_{j,n} - \tilde{\omega}) t +\psi_{j,n}  )} 
       W(\tilde{\omega} - \omega_{j,n}) ,    
\end{align*}
which completes the proof.

\section{Appendix}
\label{app:velocitydeflection}
This appendix contains a brief derivation of the relationship in equation~\cref{eq:velocitydeflection} between maximum angular velocity $\dot{\varphi}_\mathrm{max}$ and maximum deflection angle $\varphi_\mathrm{max}$. The MMRs considered in this work are 1D magneto-mechanical systems. Their rotational energy is given by
\begin{equation*}
    E_\mathrm{rot} = \frac{1}{2} I \dot{\varphi}^2,
\end{equation*}
where $I$ is the rotor's moment of inertia around the rotation axis and $\dot{\varphi}$ is its angular velocity. Assuming dipole-dipole magnetic interactions, the magnetic energy of the system, taking the equilibrium position as the reference state, is given by
\begin{equation*}
    E_\mathrm{mag} = m_\text{r}B(1-\cos\varphi),
\end{equation*}
where $m_\text{r}$ is the magnetic moment of the rotor, $B$ is the magnetic field of the stator at the rotors position, and $\varphi$ is the deflection angle from the system's equilibrium position. For the states of maximum deflection and maximum angular velocity, the total energy of the system is given by
\begin{align}
\text{I.}\quad &  E_\mathrm{rot}(\dot{\varphi}_\mathrm{max}) + 0 = \mathrm{const} \label{eq:e_rot_max} \\
\text{II.}\quad & 0 + E_\mathrm{mag}(\varphi_\mathrm{max}) = \mathrm{const}. \label{eq:e_mag_max}
\end{align}
With \cref{eq:e_rot_max} and \cref{eq:e_mag_max} the two energies relate $\varphi_\mathrm{max}$ and $\dot{\varphi}_\mathrm{max}$
\begin{equation*}
    \frac{1}{2} I \dot{\varphi}_\mathrm{max}^2 = m_\text{r}B(1-\cos\varphi_\mathrm{max}).
\end{equation*}
The natural frequency is given by $\omega_\mathrm{nat} = \sqrt{\frac{m_\text{r}B}{I}}$, as defined in the supplemental material of~\cite{gleich2023miniature}, which allows us to express the relation as
\begin{equation*}
\dot{\varphi}_\text{max} = \,\omega_\text{nat}\,\sqrt{2-2\cos(\varphi_\text{max})}
\end{equation*}
for the amplitude of the maximum angular velocity.

\section{Appendix}
\label{app:Fourierseriesderivation}
In this appendix, we derive the Fourier series \cref{Eq:phiFT} for the periodic deflection angle $\varphi^\text{TU}(t)$. According to \cite[Eq. 31]{belendez2007solNonlinearPend}, an explicit solution to \cref{Eq:DeflAngleODEUndamped} can be expressed as
\begin{equation*}
   \varphi^\text{TU}(t) = 2 \arcsin\left( k \, \text{sn}\!\left(K(k) - \frac{2K(k)}{\pi} (\omega^\text{TU}(\varphi_\text{max}) t + \psi_\text{start}); k \right) \right),
\end{equation*}
where $\text{sn}$ denotes the sinus amplitudinis, which is one of the fundamental elliptic functions \cite{Lawden1989Elliptic} and $k:= \sin (\varphi_\text{max}/2)$ denotes the elliptic modulus that we introduce to abbreviate notation. $K(k) = \int_{0}^{\pi/2} \frac{1}{\sqrt{1-k^2\sin^2 \Psi}} \,\mathrm{d}\Psi$ denotes the complete elliptic integral of the first kind. In contrast to \cite[Eq. 31]{belendez2007solNonlinearPend} we consider a non-zero initial phase and therefore added  $\psi_\text{start}$.

 Substituting $x:=\frac{2K(k)}{\pi} (\omega^\text{TU}(\varphi_\text{max}) t + \psi_\text{start})$ and using the identity $\text{sn}(x+K;k) = \text{cd}(x;k)$ \cite[Table 22.4.3]{NIST2025} leads to
\begin{equation}
   \varphi^\text{TU}(x) = 2 \arcsin\left( k \, \text{cd} \left( -x; k \right) \right).
   \label{eq:phiProof2}
\end{equation}
As the function cd is even, we can neglect the minus sign. We can then take the integral \cite[§22.14.5]{NIST2025} 
 \begin{equation}
    kk' \int \text{sd}(x';k)\,dx' = \arcsin(-k~\text{cd}(x';k)) + C
       \label{eq:phiProof3}
 \end{equation}
 with $k' := \sqrt{1-k^2}$.
 Using the property of $\arcsin$ being odd and applying the limits $K(k)$ and $ x$, we can rewrite the indefinite integral \cref{eq:phiProof3} as
 \begin{equation}
    -kk' \int^{x}_{K(k)} \text{sd}(x';k)\,dx' = \arcsin(k~\text{cd}(x;k)).
       \label{eq:phiProof4}
 \end{equation}
 Inserting \cref{eq:phiProof4} into \cref{eq:phiProof2}, we obtain 
\begin{equation}
   \varphi^\text{TU}(x) = -2 kk' \int^{x}_{K(k)} \text{sd}(x';k)\,dx'. 
   \label{eq:phiProof5}
\end{equation}
 The function $\text{sd}(x';k)$ can be written as a Fourier series using \cite[§22.11.5]{NIST2025}  

\begin{equation*}
   \text{sd}(x';k) = \frac{2\pi}{K(k)kk'} \sum_{n = 0}^{\infty} \frac{(-1)^nq(k)^{n+1/2}}{1+q(k)^{2n+1}}  \sin\left((2n+1)\frac{\pi}{2K(k)}x'\right) 
\end{equation*}
with the nome 
\begin{equation*}
q(k) := \exp \!\left( -\pi \frac{K(k')}{K (k)} \right).
\end{equation*}
Inserting this into \cref{eq:phiProof5} leads to
\begin{equation}
   \varphi^\text{TU}(x) = -\frac{4\pi}{K(k)} \sum_{n = 0}^{\infty} \frac{(-1)^nq(k)^{n+1/2}}{1+q(k)^{2n+1}} \int^{x}_{K(k)} \sin\left((2n+1)\frac{\pi}{2K(k)}x'\right)\,dx'.
   \label{eq:phiProof6}
\end{equation}
Solving the integral leads to
\begin{align}
  \int^{x}_{K(k)} \sin\left((2n+1)\frac{\pi}{2K(k)}x'\right)\,dx' &= \left[-\frac{2K(k)}{(2n+1)\pi} \cos\left( (2n+1)\frac{\pi}{2K(k)}x'\right)\right]^{x}_{K(k)} \nonumber\\
  &=-\frac{2K(k)}{(2n+1)\pi} \left( \cos\left( (2n+1)\frac{\pi}{2K(k)}x\right)-\underbrace{\cos\left( (2n+1)\frac{\pi}{2}\right)}_{=0} \right).
   \label{eq:phiProof7}
\end{align}
Inserting \cref{eq:phiProof7} into \cref{eq:phiProof6} and back-substituting $x$, we obtain 
\begin{equation}
   \varphi^\text{TU}(t) = 8 \sum_{n = 0}^{\infty} \frac{(-1)^n}{2n+1} \frac{q(k)^{n+1/2}}{1+q(k)^{2n+1}} \cos\left( (2n+1)(\omega^\text{TU}(\varphi_\text{max}) t + \psi_\text{start})\right).
   \label{eq:phiProof8}
\end{equation}
Defining 
\begin{align*}
     c_n(sk) & := \frac{(-1)^n}{2n+1}\, \frac{q(k)^{n+1/2}}{1+q(k)^{2n+1}},
\end{align*}
we can write \cref{eq:phiProof8} as
\begin{equation*}
   \varphi^\text{TU}(t) = 8 \sum_{n = 0}^{\infty} c_n(k) \cos\left( (2n+1)(\omega^\text{TU}(\varphi_\text{max}) t + \psi_\text{start})\right),
   \label{eq:phiProof9}
\end{equation*} 
which is the same as \cref{Eq:phiFT} and thus completes the derivation.

\section{Appendix}
\label{sec:appDerivationFourier}

In this appendix, we show that the induced signal can be expressed as \cref{Eq:signalUndampedExp} with amplitudes \cref{Eq:AuUndamped} and \cref{Eq:AvUndamped} when considering the undamped model with $c_n=0$ for $n>0$. Starting from \cref{Eq:SignalEquation3} and inserting the basis functions \cref{Eq:templateU} and \cref{Eq:templateV} yields
\begin{align}
    \pmb{s}^\text{TU}(t) &= m_\text{r} \pmb{\sigma}_{v} \theta_v(t)  + m_\text{r} \pmb{\sigma}_{u}\theta_u(t) = m_\text{r}\pmb{\sigma}_{v}  \dot{\varphi}(t) \cos(\varphi(t)) - m_\text{r}\pmb{\sigma}_{u} \dot{\varphi}(t) \sin(\varphi(t)).
    \label{Eq:App1}
\end{align}
We then want to eliminate the sine and cosine terms and to this end use \cref{Eq:DeflAngleODEUndamped} to replace the sine term. To substitute the cosine term, we differentiate \cref{Eq:DeflAngleODEUndamped} with respect to time, yielding
$ \dot{\varphi}(t) \cos{\varphi(t)}  = - \frac{1}{\omega_\text{nat}^2} \dddot{\varphi}(t)
$. Inserting both expressions into \cref{Eq:App1} yields
\begin{align*}
    \pmb{s}^\text{TU}(t)  & = - \frac{ m_\text{r} \pmb{\sigma}_{v}}{\omega_\text{nat}^2} \dddot{\varphi}(t) + \frac{ m_\text{r} \pmb{\sigma}_{u}}{\omega_\text{nat}^2} \dot{\varphi}(t) \ddot{\varphi}(t).
\end{align*}
Now, we can insert the series expression \cref{Eq:phiSeries} and exploit that $c_n = 0$ for $n>0$ such that
\begin{align*}  
 \pmb{s}^\text{TU}(t)  & = -\frac{8 m_\text{r} \pmb{\sigma}_{v}}{\omega_\text{nat}^2} c_0 (\omega^\text{TU})^3  \sin(\omega^\text{TU} t + \psi_\text{start})  \\
    & \quad +  \frac{64 m_\text{r} \pmb{\sigma}_{u}}{\omega_\text{nat}^2} c_0^2 (\omega^\text{TU})^3 \sin(\omega^\text{TU} t + \psi_\text{start}) \cos(\omega^\text{TU} t + \psi_\text{start}) .
\end{align*}
Note that $c_n(\varphi_\text{max})$, $\omega^\text{TU}(\varphi_\text{max})$ and $q(\varphi_\text{max})$ depend on $\varphi_\text{max}$, but for the sake of readability, we've omitted $\varphi_\text{max}$ in the equations.

Then, we exploit trigonometric expression $2\sin(x) \cos(x) = \sin(2x)$
\begin{align*}
    \pmb{s}^\text{TU}(t)   & =  -\frac{8 m_\text{r} \pmb{\sigma}_{v}}{\omega_\text{nat}^2} c_0 (\omega^\text{TU})^3  \sin(\omega^\text{TU} t+ \psi_\text{start})+ \frac{32 m_\text{r} \pmb{\sigma}_{u}}    {\omega_\text{nat}^2} c_0^2 (\omega^\text{TU})^3  \sin(2\omega^\text{TU} t + 2\psi_\text{start}).
\end{align*}

Inserting $c_0 = \frac{\sqrt{q}}{1+q}$ from \cref{Eq:phiFT} leads to
\begin{align*}    
   \pmb{s}^\text{TU}(t)  &= -\frac{8  m_\text{r}\pmb{\sigma}_{v}}{\omega_\text{nat}^2} \frac{\sqrt{q}}{1+q} (\omega^\text{TU})^3  \sin(\omega^\text{TU} t+\psi_\text{start})  + \frac{32   m_\text{r} \pmb{\sigma}_{u}}{\omega_\text{nat}^2} \frac{q}{(1+q)^2} (\omega^\text{TU})^3  \sin(2\omega^\text{TU} t+ 2\psi_\text{start}).
\end{align*} 
We then replace $\psi^\text{TU}(t) = \omega^\text{TU}(\varphi_\text{max}) t+\psi_\text{start}$ and insert the expression for $\omega^\text{TU}$ from \cref{Eq:omegaUndampedFormula} and obtain
\begin{align*}     
  \pmb{s}^\text{TU}(t)  &= \underbrace{-\frac{  \pi^3\omega_\text{nat} m_\text{r}\pmb{\sigma}_{v}}{K(\sin (\varphi_\text{max}/2))^3} \frac{\sqrt{q}}{1+q}}_{\overset{\text{\cref{Eq:AvUndamped}}}{=}\pmb{\alpha}_v(\varphi_\text{max})} \sin(\psi^\text{TU}(t))  + \underbrace{\frac{4 \pi^3\omega_\text{nat} m_\text{r}\pmb{\sigma}_{u}}{K(\sin (\varphi_\text{max}/2))^3} \frac{q}{(1+q)^2}}_{\overset{\text{\cref{Eq:AuUndamped}}}{=}\pmb{\alpha}_u(\varphi_\text{max})} \sin(2\psi^\text{TU}(t)), 
\end{align*}
which completes the proof.

\section{Additional results}
\begin{figure}[h ]
    \centering
    \includegraphics{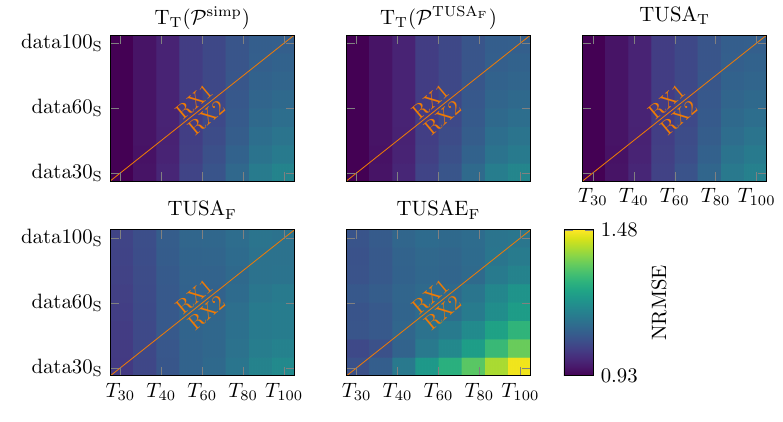}
    \caption{\textbf{Error to time signal; Experiment 1 - uncontrolled (\MMRS).} The normalized root mean square error for all methods over all frames are shown for different amounts of data and for different time ranges.}
    \label{sup_fig:errorHeatmapTimeSignal_MMRS}
\end{figure}

\begin{figure}[h ]
    \centering
    \includegraphics{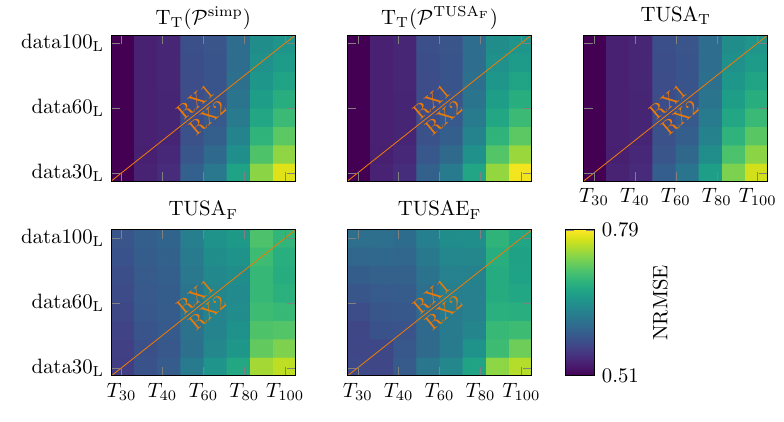}
    \caption{\textbf{Error to time signal; Experiment 1 - uncontrolled (\MMRL).} The normalized root mean square error for all methods over all frames are shown for different amounts of data and for different time ranges.}
    \label{sup_fig:errorHeatmapTimeSignal_MMRL}
\end{figure}

\begin{figure}[h]
    \centering
    \includegraphics{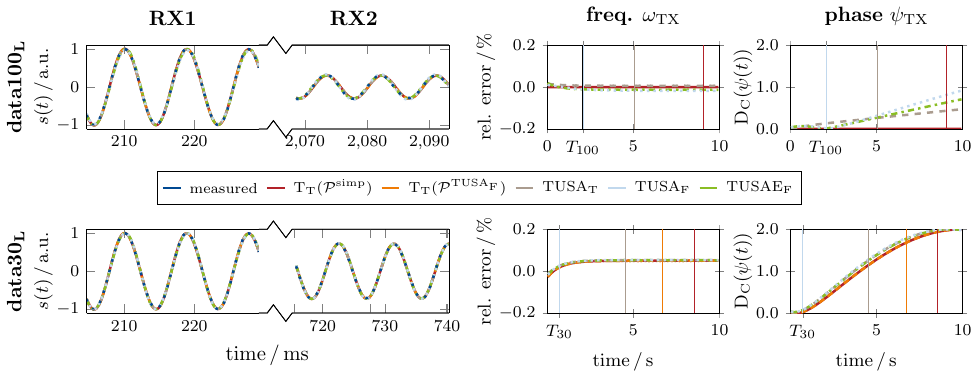}
    \caption{\textbf{Forecasts of time signal and control parameter; Experiment 1 - uncontrolled (\MMRL).} Left: Exemplary signal section at different time points of the measured signal in comparison to the estimated signals of all methods (excitation amplitude: \SI{35}{\micro T}). Signals for a time segment during RX1 as well as $\SI{100}{\milli s}$ after $T_\text{30}/T_\text{100}$ (labeled RX2) are shown. Note that all methods reconstruct the signal very closely to the measured signal within RX1. Right: The deviation for all methods from the reference method for data30$_\text{L}$ and data100$_\text{L}$ for the forecast of frequency $\omega_\text{TX}$ and phase $\psi_\text{TX}$ over time is displayed. Additionally, the cut-off time T$_\text{30}$, T$_\text{100}$ as well as the mean runtime of the methods are shown in straight lines.}
    \label{sup_fig:fittedSignals_controlParams_MMRL}
\end{figure}

\begin{figure}[t]
    \centering
    \includegraphics{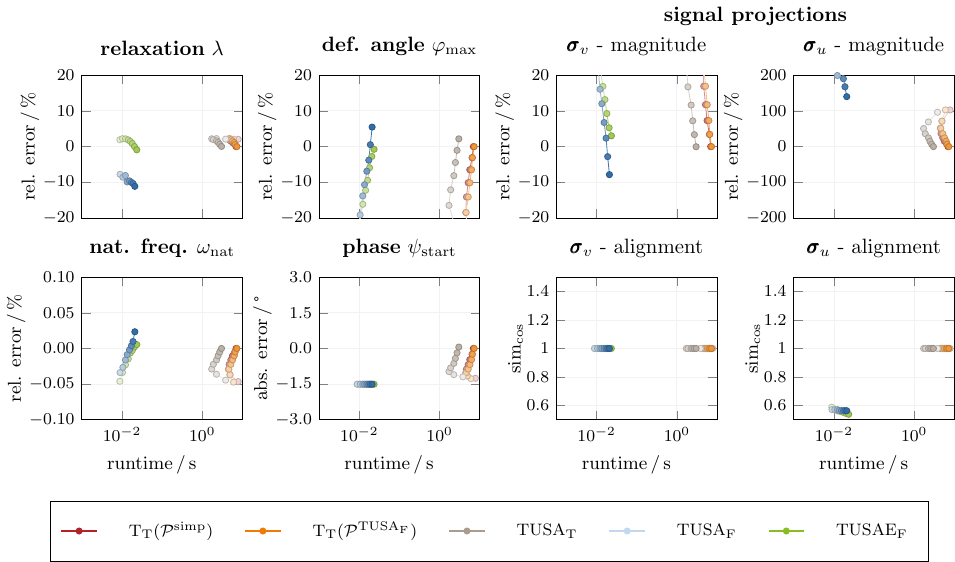}
    \caption{\textbf{Runtime and deviation for different datasets; Experiment 1 - uncontrolled (\MMRL).} The median deviation between the different methods and the reference method of all frames over the runtime for the truncated receive windows. For the parameters $\lambda,\varphi_\text{max}, \omeganat, \pmb{\sigma}_{v}$ and $\pmb{\sigma}_{u}$, the deviation is quantified using the relative error, for $\psi_\text{start}$ using the absolute error, and for the alignment of $\pmb{\sigma}_{v}$ and $\pmb{\sigma}_{u}$ using the cosine similarity. The color fading in the line plots indicates the amount of data used, with the lightest shade representing data30$_\text{L}$ and the darkest data100$_\text{L}$.}
    \label{sup_fig:MMRL_Experiment1}
\end{figure}

\begin{figure}[h]
    \centering
    \includegraphics{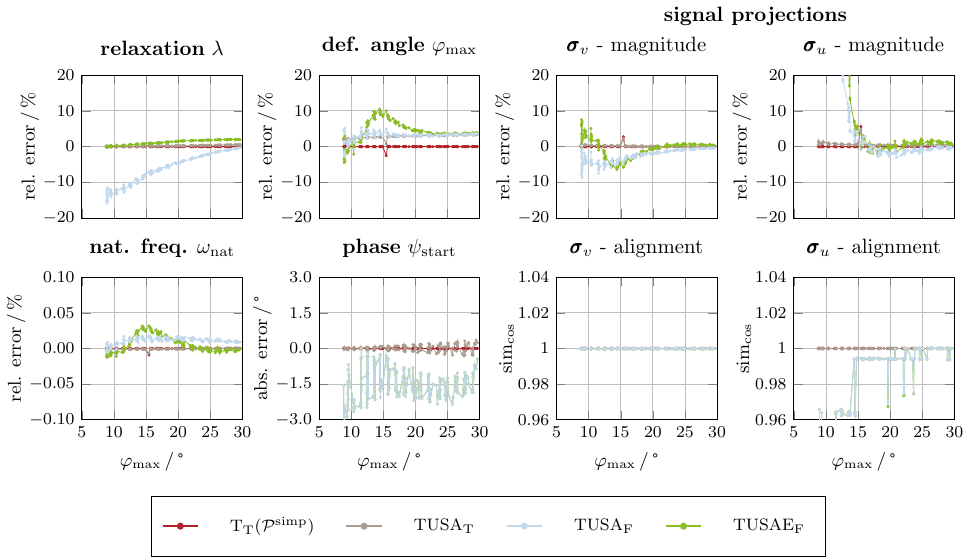}
    \caption{\textbf{Deviation for larger deflection angles; Experiment 2 - controlled (\MMRL).} The deviation of the different methods from the reference method (T$_\text{T}$$(\mathcal{P}^\text{TUSA$_\text{F}$})$) over the maximum deflection angle. For the parameters $\lambda,\varphi_\text{max}, \omeganat, \pmb{\sigma}_{v}$ and $ \pmb{\sigma}_{u}$ the deviation is quantified using the relative error, for $\psi_\text{start}$ using the absolute error, and for the alignment of $\pmb{\sigma}_{v}$ and $ \pmb{\sigma}_{u}$ using the cosine similarity.}
    \label{sup_fig:MMRL_Experiment2}
\end{figure}

\begin{figure}[h]
    \centering
    \includegraphics{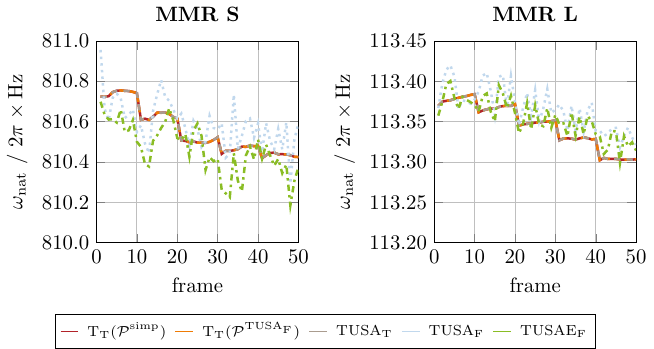}
    \caption{\textbf{Drift in estimated natural frequency; Experiment 1 - uncontrolled (\MMRS/\MMRL).} 
    The estimated natural frequency from the different methods over the measurement frame for both MMRs. Every ten frames the excitation amplitude increases in steps of \SI{0.5}{\micro T}.}
    \label{sup_fig:fnatShift}
\end{figure}

\end{document}

%% file: acronyms.tex
\usepackage[nolist,nohyperlinks]{acronym}

\begin{acronym}[ECU]
\acro{MMR}{magneto-mechanical resonator}
\acro{TX}{excitation}
\acro{RX}{receive}
\acro{S}{switch}
\acro{STFT}{short-time Fourier transform}
\acro{SP}{sensing parameter}
\acro{CP}{control parameter}
\acro{MAPE}{mean absolute percentage error}
\acro{NRMSE}{normalized root mean square error}
\acro{ODE}{ordinary differential equation}
\acro{SNR}[SNR]{signal-to-noise ratio}
\end{acronym}

%% file: math-definitions.tex
\usepackage{amsmath}
\usepackage{amsfonts}
\usepackage{amssymb}
\usepackage{nicefrac}
\usepackage{bbm} 
\usepackage{bm}  
\usepackage{nccmath} 

\usepackage{amsthm}

\usepackage[capitalise,nameinlink]{cleveref} 
\crefformat{equation}{(#2#1#3)}

\newcommand{\IR}{\mathbb{R}}

\renewcommand{\epsilon}{\varepsilon}				
\renewcommand{\theta}{\vartheta}                  	
\renewcommand{\phi}{\varphi}						

\newcommand{\set}[2][]{\left\{ #2 
                        \ifthenelse{\equal{#1}{}}{}{: #1}
                        \right\}}				    

\newcommand{\MMRL}{MMR~L}       
\newcommand{\MMRS}{MMR~S}         
\newcommand{\omeganat}{\omega_\text{nat}}

\usepackage[exponent-product = \cdot, binary-units]{siunitx}
\DeclareSIUnit{\mup}{\text{$\mu_0$}}
\DeclareSIUnit{\Tpm}{\tesla\per\meter} 
\DeclareSIUnit{\T}{\tesla} 

\usepackage{siunitx}
\usepackage{tikz}
\usepackage{tikz-3dplot}
\usepackage{pgfplots}
\usepackage{pgfplotstable}
\usepgfplotslibrary{groupplots} 
\pgfplotsset{compat=1.8}

\usetikzlibrary{shapes,shapes.multipart,positioning,fit}
\usetikzlibrary{arrows.meta, 3d,  perspective, calc}
\usepgfplotslibrary{units,statistics}
\usetikzlibrary{backgrounds} 

\RequirePackage{xcolor}
\definecolor{ibilight}{RGB}{193,216,237}
\definecolor{ibidark}{RGB}{0,73,146}	
\definecolor{uke2}{RGB}{170,156,143} 	
\definecolor{uke3}{RGB}{87,87,86}		
\definecolor{ukesec1}{RGB}{255,223,0}	
\definecolor{ukesec2}{RGB}{239,123,5}	
\definecolor{ukesec3}{RGB}{104,195,205}
\definecolor{ukesec4}{RGB}{138,189,36}	
\definecolor{ukesec5}{RGB}{178,34,41}	
\definecolor{tuhh}{RGB}{45,198,214}    

\definecolor{ibidarkBG}{RGB}{227,229,242}  
\definecolor{uke2BG}{RGB}{233,228,225} 	  
\definecolor{uke3BG}{RGB}{230,231,232}	   
\definecolor{ukesec1BG}{RGB}{255,243,190}  
\definecolor{ukesec2BG}{RGB}{254,232,212}  
\definecolor{ukesec3BG}{RGB}{222,241,241}  
\definecolor{ukesec4BG}{RGB}{233,243,222}  
\definecolor{ukesec5BG}{RGB}{244,230,225}  

\pgfplotsset{
  colormap={colorModelAOne}{
    color(0cm)=(ukesec5!20);
    color(1cm)=(ukesec5!80)
  }
}
\pgfplotsset{
  colormap={colorModelATwo}{
    color(0cm)=(ukesec2!20);
    color(1cm)=(ukesec2!80)
  }
}
\pgfplotsset{
  colormap={colorModelB}{
    color(0cm)=(uke2!20);
    color(1cm)=(uke2!80)
  }
}
\pgfplotsset{
  colormap={colorModelD}{
    color(0cm)=(ukesec4!20);
    color(1cm)=(ukesec4!80)
  }
}
\pgfplotsset{
  colormap={colorModelE}{
    color(0cm)=(ibidark!20);
    color(1cm)=(ibidark!80)
  }
}